\documentclass[sigconf]{acmart}
\usepackage{algorithmic}
\usepackage{multirow}
\usepackage{adjustbox}
\usepackage[acronym]{glossaries}
\usepackage{url}
\usepackage{hyperref}

\newacronym{iou}{IOU}{intersection over union}
\newacronym{tp}{TP}{true positive}
\newacronym{fp}{FP}{false positive}
\newacronym{fn}{FN}{false negative}
\newacronym{tn}{TN}{true negative}
\newacronym{fpn}{FPN}{feature pyramid network}
\newacronym{gi}{GI}{gastrointestinal}
\newacronym{ai}{AI}{artificial intelligence}
\newacronym{gan}{GAN}{generative adversarial network}
\newacronym{cad}{CAD}{computer-aided diagnosis}
\newacronym{fid}{FID}{Fréchet inception distance}
\newacronym{sim}{SIM}{similarity measure}
\newacronym{dpm}{DPM}{diffusion probabilistic model}
\newacronym{ml}{ML}{machine learning}
\newacronym{gpu}{GPU}{graphic processing unit}

\AtBeginDocument{%
  }

\setcopyright{acmcopyright}
\copyrightyear{2018}
\acmYear{2018}
\acmDOI{XXXXXXX.XXXXXXX}

\acmConference[Conference acronym 'XX]{Make sure to enter the correct
  conference title from your rights confirmation emai}{June 03--05,
  2018}{Woodstock, NY}
\acmPrice{15.00}
\acmISBN{978-1-4503-XXXX-X/18/06}

\begin{document}

\title{Mask-conditioned latent diffusion for generating gastrointestinal polyp  images}


\author{Roman Macháček}
\authornote{Contributed equally to this research.}
\affiliation{%
  \institution{University of Oslo}
  \city{Oslo}
  \country{Norway}}
\email{ro.machacek0@gmail.com}

\author{Leila Mozaffari}
\authornotemark[1]
\affiliation{%
  \institution{Oslo Metropolitan University}
  \city{Oslo}
  \country{Norway}}
\email{s372060@oslomet.no}

\author{Zahra Sepasdar}
\affiliation{%
  \institution{Monash University}
  \city{Melbourne}
  \country{Australia}}
\email{zahra.sepasdar@monash.edu}

\author{Sravanthi Parasa}
\affiliation{%
  \institution{Swedish Medical Group}
  \city{Seattle}
  \country{USA}}
\email{vaidhya209@gmail.com}

\author{P{\aa}l Halvorsen}
\email{paalh@simula.no}
\orcid{1234-5678-9012}
\affiliation{%
  \institution{SimulaMet}
  \city{Oslo}
  \country{Norway}}
\authornote{Also affiliated with Oslo Metropolitan University, Norway.}

\author{Michael A. Riegler}
\authornotemark[2]
\email{michael@simula.no}
\orcid{1234-5678-9012}
\affiliation{%
  \institution{SimulaMet}
  \city{Oslo}
  \country{Norway}}

\author{Vajira Thambawita}
\authornotemark[1]
\orcid{1234-5678-9012}
\email{vajira@simula.no}
\affiliation{%
  \institution{SimulaMet}
  \city{Oslo}
  \country{Norway}}

\renewcommand{\shortauthors}{Roman et al.}

\begin{abstract}
In order to take advantage of \gls{ai} solutions in endoscopy diagnostics, we must overcome the issue of limited annotations. These limitations are caused by the high privacy concerns in the medical field and the requirement of getting aid from experts for the time-consuming and costly medical data annotation process.  In computer vision, image synthesis has made a significant contribution in recent years, as a result of the progress of \glspl{gan} and \glspl{dpm}. Novel DPMs have outperformed GANs in text, image, and video generation tasks. Therefore, this study proposes a conditional \gls{dpm} framework to generate synthetic \gls{gi} polyp images conditioned on given generated segmentation masks. Our experimental results show that our system can generate  an unlimited number of high-fidelity synthetic polyp images with the corresponding ground truth masks of polyps.  To test the usefulness of the generated data we trained binary image segmentation models to study the effect of using synthetic data. Results show that the best micro-imagewise \gls{iou} of $0.7751$ was achieved from DeepLabv3+ when the training data consists of both real data and  synthetic data. However, the results reflect that achieving good segmentation performance with synthetic data heavily depends on model architectures. 
\end{abstract}



\received{20 February 2007}
\received[revised]{12 March 2009}
\received[accepted]{5 June 2009}

\maketitle
\glsresetall
\section{Introduction}

The human digestive system can experience a range of abnormal tissue changes, from minor discomforts to severe, life-threatening illnesses \cite{kaminski2010quality}. Endoscopy, colonoscopy, and pilcams (wireless capsule endoscopy)~\cite{iddan2000wireless} are the most common methods for examining the \gls{gi} tract for diagnosis. However, its effectiveness is greatly impacted by the variability in the performance of the operator (inter-rater reliability) \cite{borgli2020hyperkvasir}. In this regard, \gls{ai} techniques are researched to build \gls{cad} systems to aid gastroenterologists~\cite{le2020application, riegler2016eir, vinsard2019quality, thambawita2020extensive}.

Supervised machine learning models have become popular in many applications, such as image classification, image detection, and image segmentation. However, \gls{ai} models require vast amounts of data to train. Especially supervised machine learning techniques need annotated datasets to train. In  the medical field, however, acquiring a large annotated dataset is challenging. The challenges include not only privacy concerns but also costly and timely medical data labeling and annotation. In comparison with other applications of machine learning in the health area, we have limited annotated datasets to train \gls{ml} models. \Gls{gi}-tract  datasets are also typically small and mostly limited to polyps \cite{min2019computer}. To overcome this issue, one solution is to expand training datasets by generating synthetic data \cite{jordon2022synthetic, thambawita2021deepsynthbody}.
 
Endoscopy diagnostics are being enhanced by \gls{ai} solutions. Especially, image synthesis has made a significant contribution to overcome the issue of the limited dataset  \cite{rombach2022high}. 
It is now common to use \glspl{gan} to generate synthetic images because \glspl{gan} produce realistic images and achieve impressive results in a wide range of applications \cite{creswell2018generative, alqahtani2021applications}. Thus, a \gls{gan} is a powerful generative model, however, it suffers from convergence instability. 

To overcome the convergence issue in \glspl{gan}, in recent years, diffusion models~\cite{ho2020denoising} have gained attention as a potential method for their ability to synthesize natural images.
In this study, we introduce a framework consisting of two different diffusion models to create synthetic \gls{gi}-tract images and corresponding masks. Our contributions are listed as follows:
\begin{itemize}
    \item We introduce a fully synthetic polyp generation system.
    \item Our system is able to generate realistic-looking synthetic polyp masks using an improved diffusion model. 
    \item Based on the generated masks, we are able to generate high-fidelity synthetic polyp images conditioned on pre-generated synthetic polyp masks using a conditional latent diffusion model. 
    \item We provide a comprehensive evaluation of using synthetic polyp and mask data to train polyp segmentation models and overall results. 
\end{itemize}

The source code of all the experiments is available at \url{https://github.com/simulamet-host/conditional-polyp-diffusion} and the pre-generated synthetic masks and the corresponding conditional synthetic polyp images are available at \url{https://huggingface.co/datasets/deepsynthbody/conditional-polyp-diffusion }. 

\section{Related work}


There are many \gls{gi} image analysis datasets available for machine learning tasks. Some of the commonly used datasets in human \gls{gi} tract are: ETIS-Larib \cite{silva2014toward}, CVC-ClinicDB \cite{bernal2015wm}, ASU-Mayo Clinic Polyp database \cite{tajbakhsh2015automated}, Kvasir \cite{pogorelov2017kvasir}, Kvasir-SEG \cite{jha2020kvasir} and Hyperkvasir~\cite{borgli2020hyperkvasir}.
A few datasets containing manually annotated segmentation masks for polyps. However, these real-world datasets (not limited to \gls{gi}-tract data) have some limitations. The limitations include: 

\begin{itemize}
    \item Size: medical image datasets, including those for polyp detection and segmentation, are often smaller in size compared to other image datasets, such as ImageNet~\cite{krizhevsky2017imagenet}, Microsoft COCO~\cite{lin2014microsoft} which can limit their ability to train complex machine learning models.
     \item Annotation quality: the accuracy and consistency of the annotations of the dataset can impact the performance of machine learning algorithms. Annotations are dependent on annotator and normally high inter-rater variability is there. 
     \item Diversity: the diversity of the images in the dataset is important for the generalization of machine learning algorithms. If the dataset is limited to a narrow range of images, the algorithm may not perform well on new, unseen images.
     \item Accessibility: legal and privacy constraints can limit the accessibility of medical image datasets, making it difficult to obtain large and diverse datasets for machine learning tasks.
\end{itemize}

These limitations highlight the need for ongoing development and improvement of medical image datasets to support the advancement of machine learning in medical imaging. To overcome these limitations of real-world datasets, synthetic datasets\cite{9462062, thambawita2021deepfake, thambawita2022singan, thambawita2021id, chen2021synthetic} have been increasingly used in medical image analysis. For instance, to generate synthetic polyps, a \gls{gan} framework has been proposed to generate a polyp mask and then synthesize the generated polyp with the real polyp image without the use of additional datasets and processes~\cite{qadir2022simple}. There has also been research on the augmenting of colonoscopy images with polyps by using synthetic samples~\cite{adjei2022examining}. Fagereng et al.~\cite{fagereng2022polypconnect} present a solution to overcome the challenge of a lack of annotated data when building \gls{cad} systems for detecting polyps in the \gls{gi}-tract. The authors propose a pipeline called PolypConnect, which can convert non-polyp images into polyp images to increase the size of training datasets for machine learning models. The results show $5.1\%$ improvement in mean \gls{iou} when using synthetic data in addition to real data for training. Dhariwal et al.~\cite{dhariwal2021diffusion} compare the performance of diffusion models and \glspl{gan} for image synthesis tasks. As a result, the authors found that diffusion models outperformed \glspl{gan} in terms of image quality and stability of the generated images. The results of the paper indicate that diffusion models are a promising alternative to \glspl{gan} in the field of image synthesis. This work provides valuable insights into the strengths and limitations of both diffusion models and \glspl{gan}.

\section{Methodology}
\begin{figure}[htbp]
\centerline{\includegraphics[width=9cm, height=9cm]{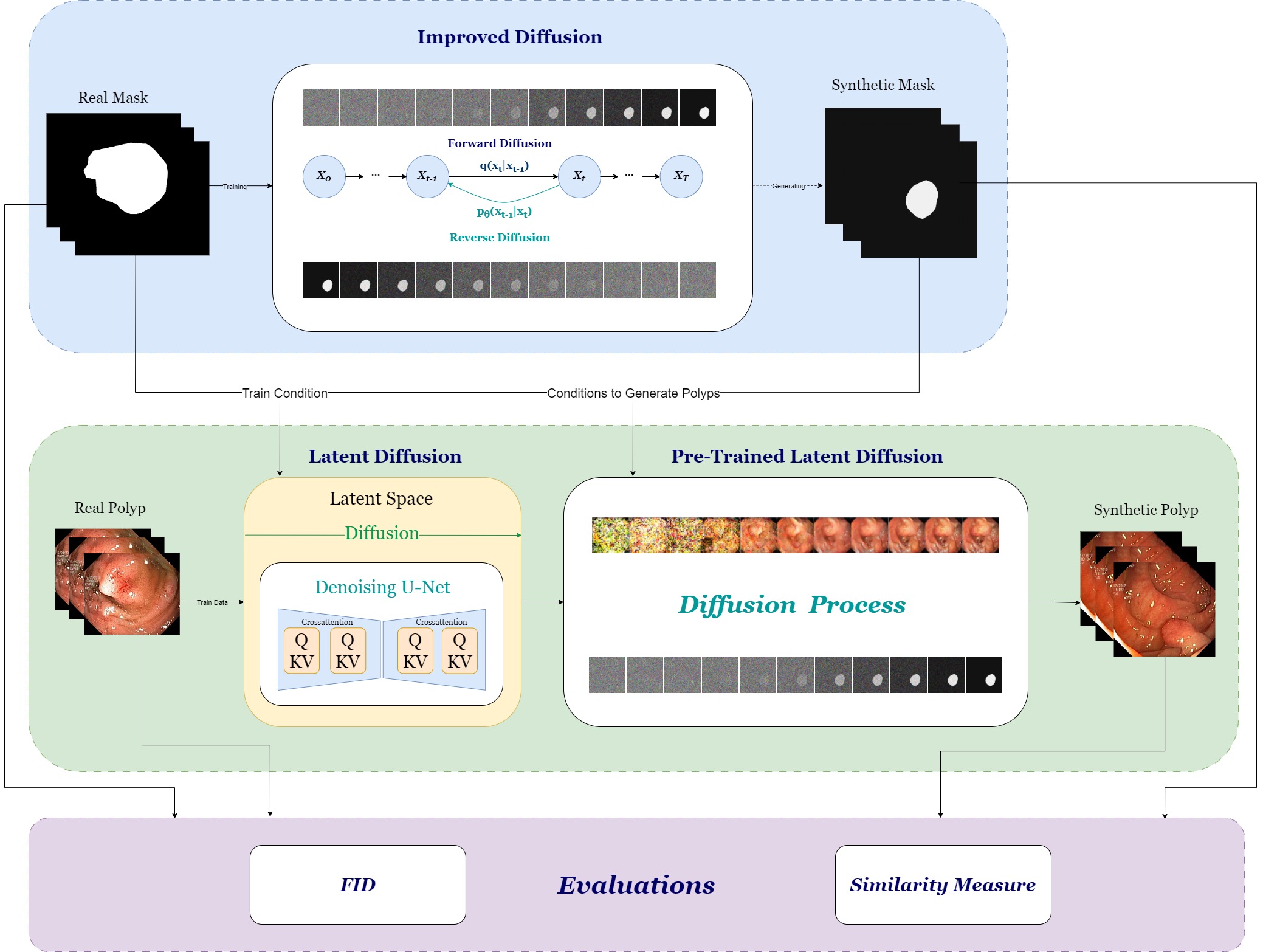}}
\caption{The whole pipeline of generating synthetic polyps and mask. The blue box represents the diffusion model trained to generate realistic synthetic polyp masks. The green box represents the conditional latent diffusion model which is used to generate synthetic polyp conditioned on input masks. The bottom box represents the evaluation matrices.}
\label{Pipline_Diffusion_Model}
\end{figure}

In this section, first, we describe the improved diffusion model which is used to generate realistic synthetic polyp mask images (the blue box of Figure~\ref{Pipline_Diffusion_Model}). Then, we present the latent diffusion model which is used for generating synthetic polyp images conditioned on the input masks generated from our aforementioned mask generator (the green box of Figure~\ref{Pipline_Diffusion_Model}).  We evaluate the quality of the generated synthetic data and quantify similarity between generated and real data, representing using the last section of Figure~\ref{Pipline_Diffusion_Model}. Finally, we present our methods used to check the quality of synthetic data using image segmentation models.  

\subsection{Improved diffusion model}

In our pipeline, we use an \emph{improved diffusion model}~\cite{nichol2021improved} to generate synthetic polyp masks which looks realistic to capture the distribution of the masks of the Kvasir-SEG dataset. The \emph{improved diffusion model} is a type of generative model that uses a gradual, multi-step process to match a data distribution and generate synthetic images. In the context of generating synthetic mask images for the \gls{gi}-tract, improved diffusion models can be used to generate synthetic  mask images that closely resemble real images. Therefore, these models overcome the issue of limited annotated data \cite{nichol2021improved,sohl2015deep}.

To achieve our goal, the first step is to obtain a training set for real mask images of the \gls{gi} tract indicating the location of polyps. Then, the mask dataset is used to train the improved diffusion model to generate synthetic polyp images that closely resemble the real images. The improved diffusion model generates synthetic mask images by first adding noise to a randomly selected mask image from the training set. This noise would then be gradually reversed through multiple steps until a synthetic mask image is generated.

The advantage of using \emph{improved diffusion models} to generate synthetic
images is that we can overcome the issue of limited annotated data and train machine learning models more effectively. This
can lead on to improvements in the accuracy and efficiency of
\gls{cad} systems for detecting polyps
in the \gls{gi}-tract \cite{fagereng2022polypconnect}

\subsection{Latent diffusion model}

The Latent Diffusion model~\cite{rombach2022high}, developed by CompVis and trained on the LAION-400M dataset~\cite{schuhmann2021laion}, operates through a series of denoising autoencoders and diffusion models. This model has been utilized to generate images based on text prompts, and has shown exceptional results in tasks related to image inpainting~\cite{elharrouss2020image} and various other applications, surpassing the current state of the art~\cite{rombach2021highresolution}.

Latent diffusion models are a suitable choice for generating synthetic images of the \gls{gi} tract for several reasons. Firstly, they possess the ability to model intricate and non-linear patterns in the data, crucial for producing convincing images of the \gls{gi} tract. Secondly, they are capable of generating a large diversity of high-quality synthetic images, which enhances the generalizability of machine learning models. Lastly, they can be trained with a limited amount of real data points, which is important in medical imaging where annotated data is often scarce.

\subsection{Mask similarity}
To assess the quality of generated images for generative models, the \gls{fid}~\cite{heusel2017gans} metric is typically used, which compares the distribution of the generated images compared to real images. Because the improved diffusion model are generating binary masks of polyps, we can also introduce the \gls{sim} metric that is analogous to accuracy. Consider real image $r$ together with generated image $g$ of the same size, then the similarity $sim$ is defined as number of pixels that are same for both images divided by the total area of image:

$$sim(r, g) = \dfrac{\#(r == g)}{width * height}$$

To measure the similarity $SIM$ of generated images to our training real images $R$, we simply take average of the closest pairs (high similarities). If a generated image is $g$ and associated closest real image is $g*$, we can calculate largest similarity using,

$$SIM(R, G) = \dfrac{1}{|G|}\sum_{r \in R}{sim(r, g*)}$$

The idea here is that even if the generated images are highly similar to some of the training images (same size, position), they should differ in another aspects, such as rotation.

\subsection{Segmentation models}

We have used three different well-known image segmentation models, namely UNet++~\cite{zhou2018unet++}, \gls{fpn}~\cite{lin2017feature}, and DeepLabv3+~\cite{chen2017rethinking} for evaluating the effect of using synthetic data for training polyp segmentation tasks. Initially, we trained these three models using three different approaches, i.e., we trained the system using i) $700$ of real polyp images; ii) using $1000$ synthetic polyp images; and iii) a combination of $700$ real and $1000$ synthetic polyp images. 
To further analyze the effect of synthetic data, we trained these three models with another set of real and synthetic data combinations. In these combinations, we fixed the number of real images to $100$ samples, and we increased the number of synthetic samples from $0$ to $1000$ sequentially in steps of $100$. The main objective of this experiment is to identify the effect of a number of synthetic samples included in the training data. However, we limited this experiment to using only $100$ real images because of the time limitation, but in the future we will test with different number of real images from $200$ to $700$ to find the optimal combination to get better performance.    

We tested these models with $300$ real images and masks (from the segmentation data of HyperKvasir dataset~\cite{borgli2020hyperkvasir}) which were not used to train either the diffusion model or the segmentation models. Then, we measured micro and micro-imagewise  \gls{iou}, F1, Accuracy, and Precision from the test dataset for all the segmentation models. Micro values were calculated by summing \gls{tp}, \gls{fp}, \gls{fn}, and \gls{tn} pixels over all images and all classes and then computing scores. In contrast, the micro-imagewise matrices were calculated by summing \gls{tp}, \gls{fp}, \gls{fn}, and \gls{tn} pixels for each image and then computing scores for each image. Finally, average scores over the dataset were calculated. In the micro-imagewise calculations, all images contributed equally to the final score. However, the second method takes into account class imbalance for each image.

\section{Results and discussion}

In this section, we discuss experiment setup and the result collected from generative models and segmentation models. A server with Nvidia A100 $80GB$ \glspl{gpu}, AMD EPYC 7763 64-cores processor with $2TB$ RAM were used for all the experiments of this study. Additionally, we used Pytorch~\cite{NEURIPS2019_9015}, the Pytorch-lightning libraries, and the Pytorch segmentation library~\cite{Iakubovskii:2019} as development frameworks. 

\subsection{Diffusion experiments and results}

For mask generator, that is improved diffusion model we have used \gls{fid} and \gls{sim} values to quantify and select appropriate model. We have generated $1000$ masks for each of our saved model, and we compare them with $1000$ real training masks in Table \ref{tab:mask}.

\begin{table}[!t]
\centering
    \begin{tabular}{ccccccc} \toprule
        \textbf{Iter} & 0  & 50k & 100k & 150k & 200k & 230k   \\ \midrule
        \textbf{FID} & 140.14  & 128.95 & 117.14 & 105.63 & 88.41 & 141.44 \\ 
        \textbf{SIM} & 88.22  & 89.46 & 90.81 & 91.31 & 92.49 & 88.38 \\\bottomrule
    \end{tabular}
    \caption{Comparison of mask models based on FID, SIM}
    \label{tab:mask}
\end{table}

We selected the model from iteration $200,000$ based on the results from Table \ref{tab:mask}. The reason is that the model achieves lowest \gls{fid} value together with high \gls{sim} values, indicating diverse and quality masks. We also inspected the generated masks visually to confirm this conclusion. 

\begin{figure*}[!t]
        \centering
        \setlength{\tabcolsep}{1pt}
        \newcommand{\imgwidth}{0.18\linewidth}
        \begin{tabular}{cccccc}
        \multirow{-8}{*}{Real} &
        \includegraphics[width=\imgwidth,height=\imgwidth]{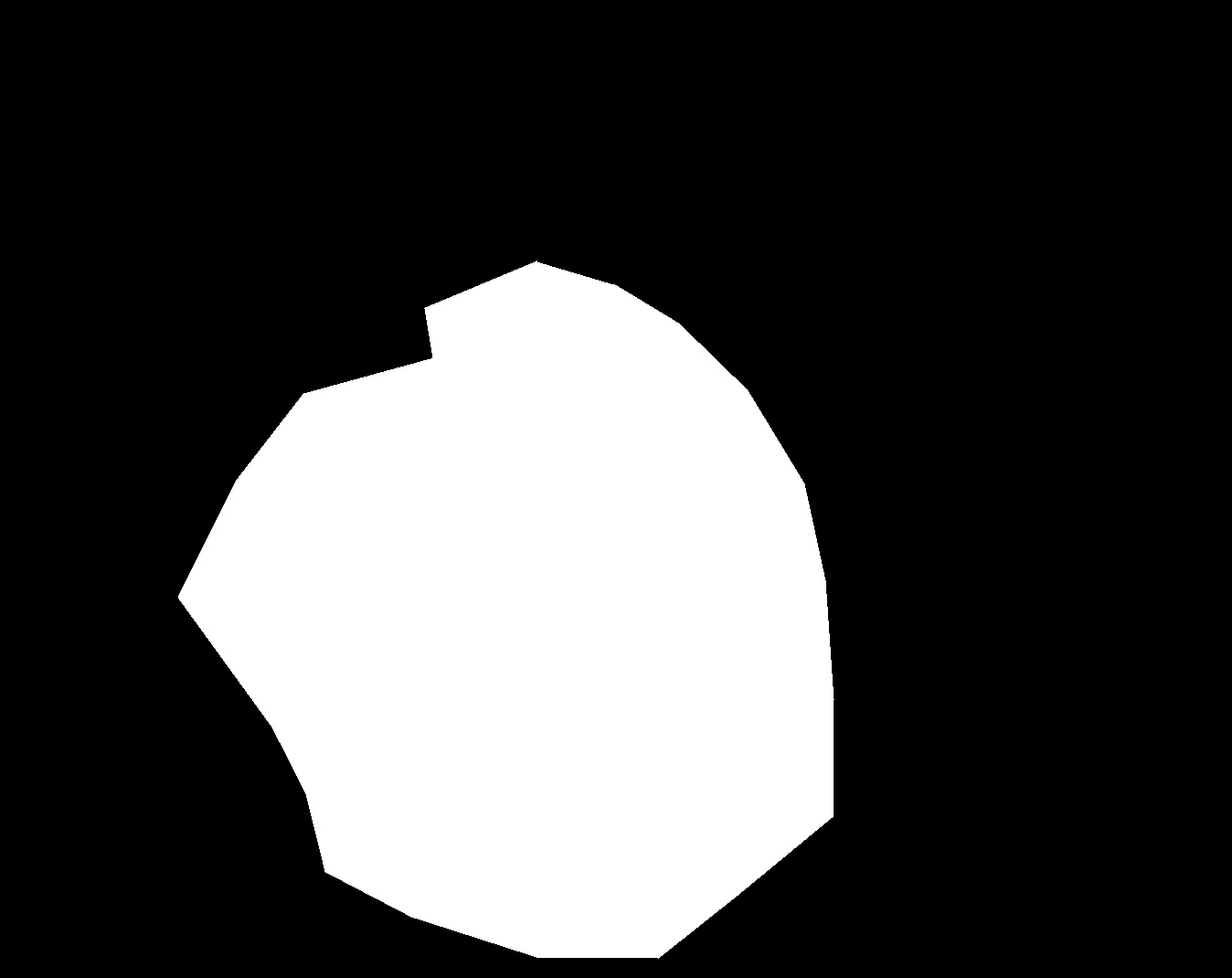} &
        \includegraphics[width=\imgwidth,height=\imgwidth]{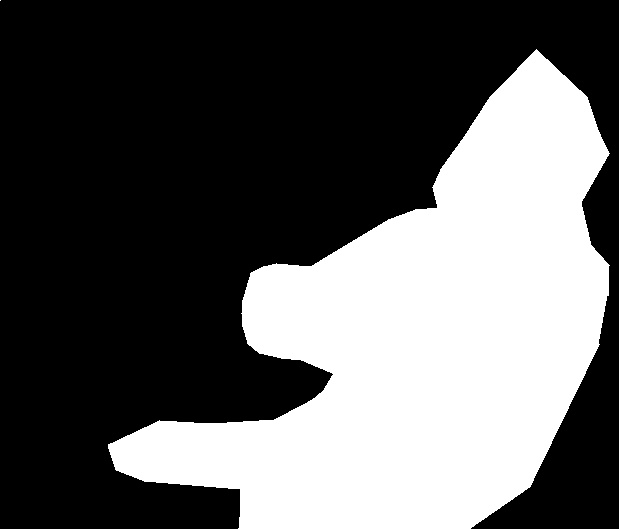} &
        \includegraphics[width=\imgwidth,height=\imgwidth]{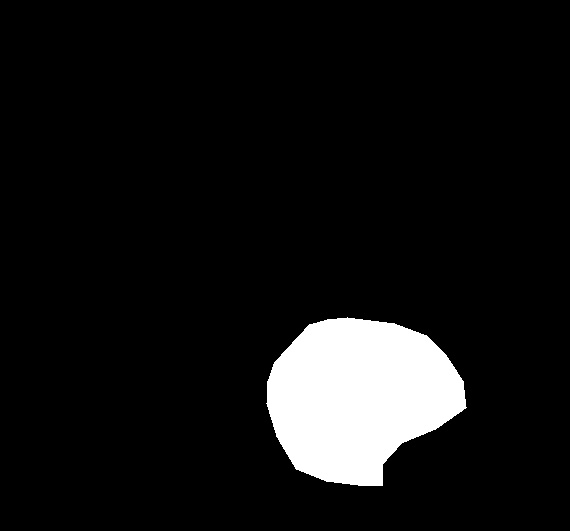} &
        \includegraphics[width=\imgwidth,height=\imgwidth]{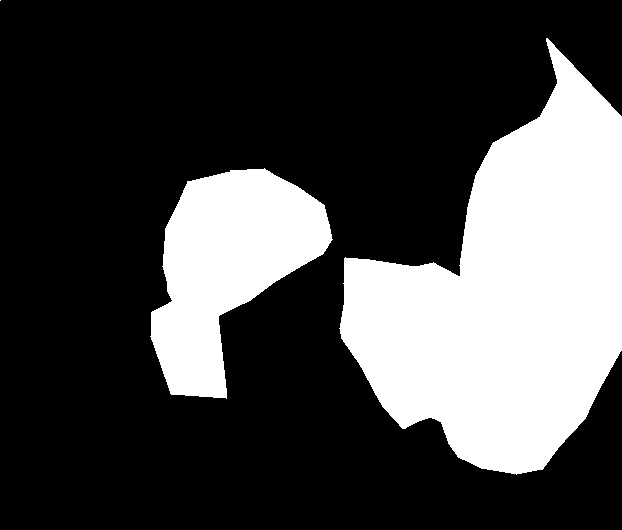} &
        \includegraphics[width=\imgwidth,height=\imgwidth]{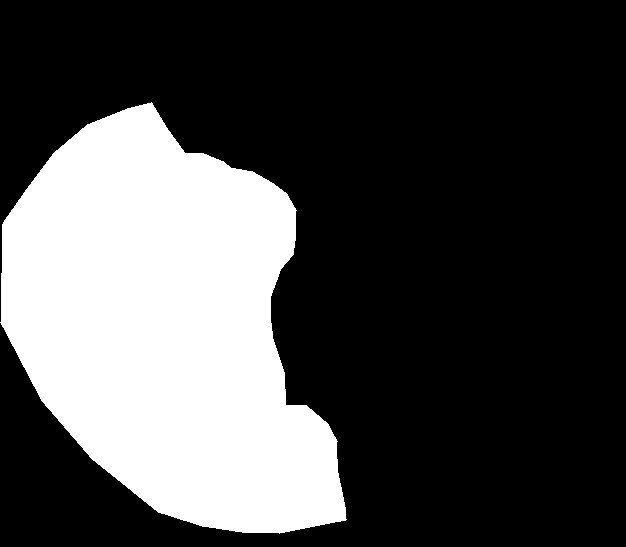} 
        \\
        \multirow{-8}{*}{Generated} &
        \includegraphics[width=\imgwidth]{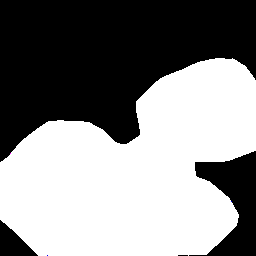} &
        \includegraphics[width=\imgwidth]{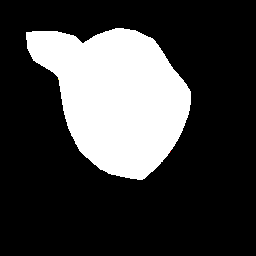} &
        \includegraphics[width=\imgwidth]{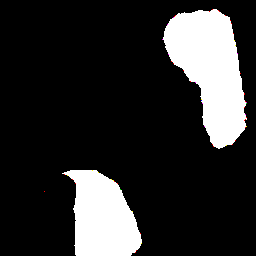} &
        \includegraphics[width=\imgwidth]{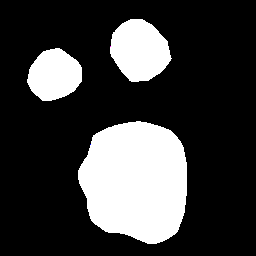} &
        \includegraphics[width=\imgwidth]{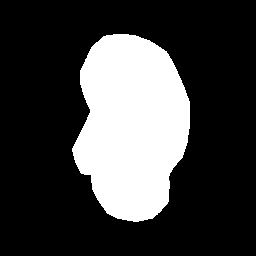} 
        \end{tabular}
    \caption{Examples of real masks in the first row, generated masks on the second. Note the variability of shapes and amount of polyps in the generated masks.}
    \label{fig:generated_masks}
\end{figure*}

Examples of generated masks in comparison with real masks can be seen in Figure \ref{fig:generated_masks}. We can see different masks with different shapes and numer of polyps indicating capability to generate diverse synthetic masks. Further discussion with medical professional would be required in order to determine if masks are correct.

\begin{figure*}[!th]
        \centering
        \setlength{\tabcolsep}{1pt}
        \newcommand{\imgwidth}{0.18\linewidth}
        \begin{tabular}{cccccc}
        SIM 
        & 95.89 & 98.11 & 93.45 & 84.62 & 87.98 \\ 
        \multirow{-8}{*}{Real} 
        & \includegraphics[width=\imgwidth, 
        height=\imgwidth]{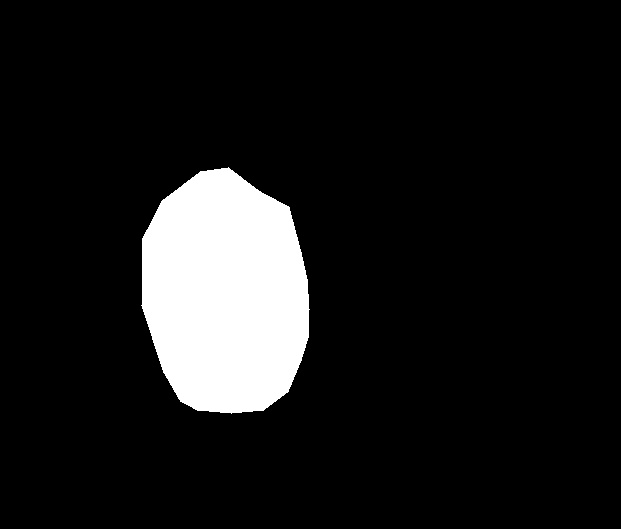} & \includegraphics[width=\imgwidth, 
        height=\imgwidth]{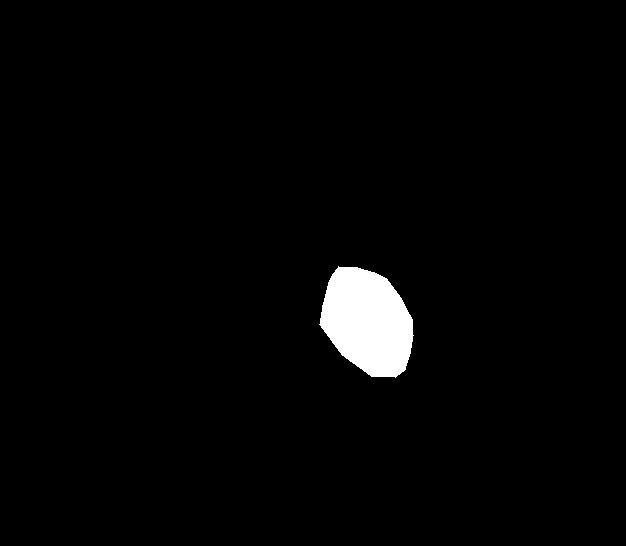} & \includegraphics[width=\imgwidth, 
        height=\imgwidth]{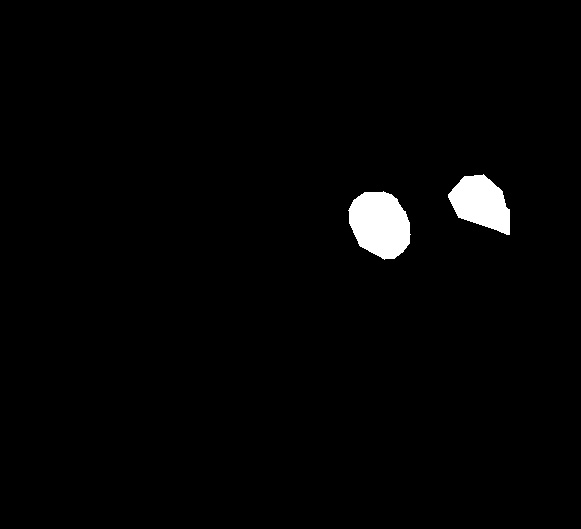} & \includegraphics[width=\imgwidth, 
        height=\imgwidth]{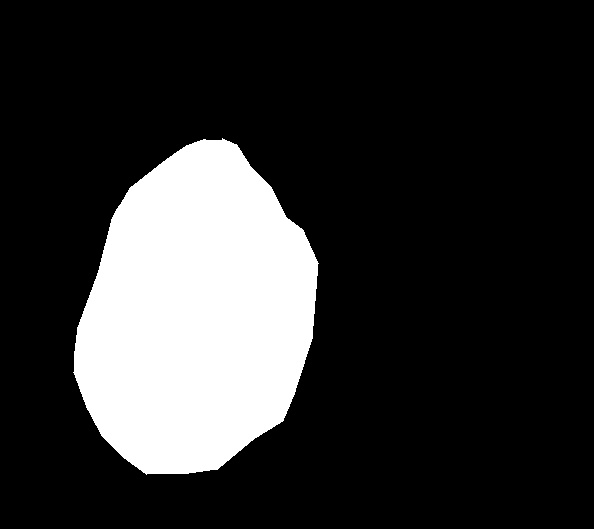} & \includegraphics[width=\imgwidth, 
        height=\imgwidth]{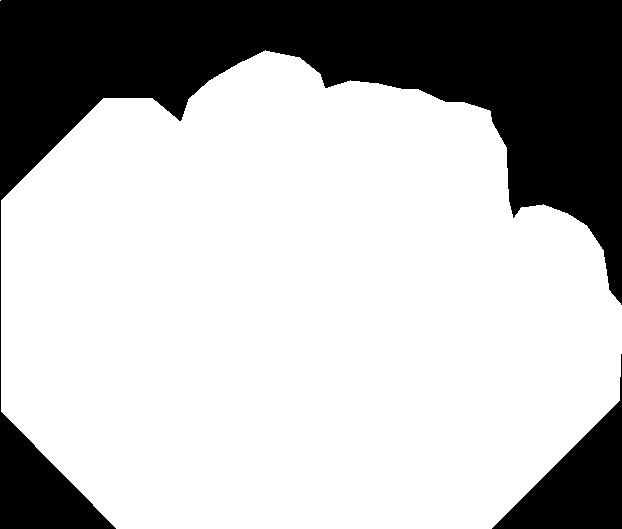} \\
        \multirow{-8}{*}{Generated}
        & 
        \includegraphics[width=\imgwidth, 
        height=\imgwidth]{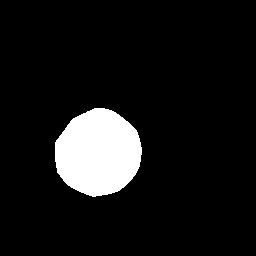} & \includegraphics[width=\imgwidth, 
        height=\imgwidth]{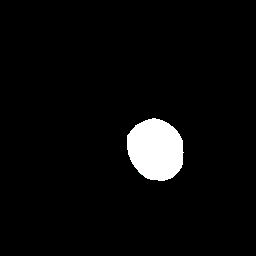} & \includegraphics[width=\imgwidth, 
        height=\imgwidth]{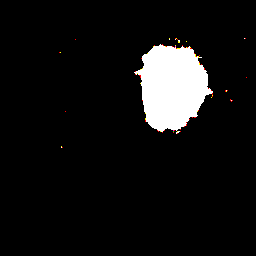} & \includegraphics[width=\imgwidth, 
        height=\imgwidth]{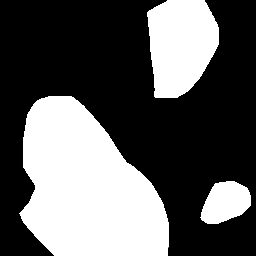} & \includegraphics[width=\imgwidth, 
        height=\imgwidth]{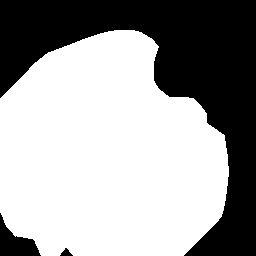} \\
        \end{tabular}
  \caption{Examples of comparison of generated masks $g$ to real masks $r$ based on similarity measure $sim(r, g)$.}
  \label{fig:similarities}
\end{figure*}

Interestingly, high $SIM$ score doesn't necessary imply that model is producing identical masks, as can be seen in the Figure \ref{fig:similarities}. For instance, masks may be located in similar positions but have different, smaller shapes, therefore achieving higher similarities.

We have used the generated masks made by the selected model as conditions to our latent polyp diffusion model and produced $1000$ generated images which we used for further evaluation in Table \ref{tab:polyps}.

\begin{table}[!t]
\centering
    \begin{tabular}{ccccccc} \toprule
        \textbf{Epoch} & 88  & 103 & 135 & 892 & 913 & 922   \\ \midrule
        \textbf{FID} & 119.34  & 113.83 & 104.78 & 112.66 & 150.97 & 150.85 \\ \bottomrule
    \end{tabular}
    \caption{Comparison of polyp models based on FID}
    \label{tab:polyps}
\end{table}

\begin{figure*}[!t]
        \centering
        \setlength{\tabcolsep}{1pt}
        \newcommand{\imgwidth}{0.13\linewidth}
        \begin{tabular}{cccccccc}
        Condition & Epoch=88 & Epoch=103 & Epoch=135 & Epoch=892 & Epoch=913 & Epoch=922 \\
        \includegraphics[width=\imgwidth,height=\imgwidth]{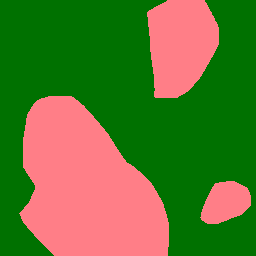} &
        \includegraphics[width=\imgwidth,height=\imgwidth]{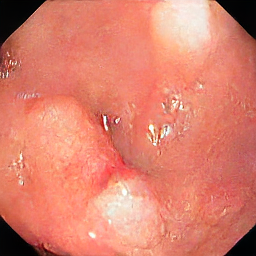} &
        \includegraphics[width=\imgwidth,height=\imgwidth]{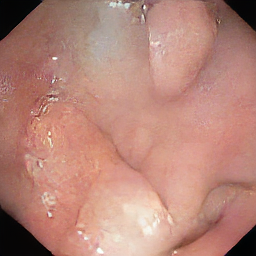} &
        \includegraphics[width=\imgwidth,height=\imgwidth]{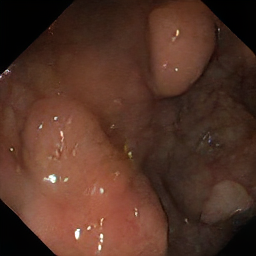} &
        \includegraphics[width=\imgwidth,height=\imgwidth]{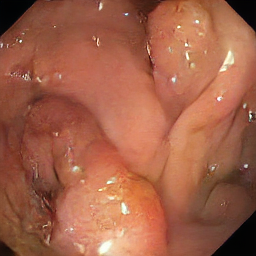} & 
        \includegraphics[width=\imgwidth,height=\imgwidth]{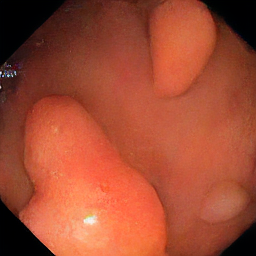}
        & 
        \includegraphics[width=\imgwidth,height=\imgwidth]{Images/polyps_progress/samples_1000x256x256x3_4_post_4.png}
        \end{tabular}
  \caption{Generated synthetic polyps conditioned on the same mask illustrating changes in quality during training stages.}
    \label{fig:polyps_progress}
\end{figure*}

We can see from Table \ref{tab:polyps} the model which achieved lowest \gls{fid} score is at $Epoch=135$. We inspected the generated images, similarly as in Figure \ref{fig:polyps_progress}. It can be seen that the quality of generated images deteriorates at later stages of training, reason may be overfitting. This may lead to problems while generating different samples with same condition which would be more similar.

\begin{figure*}[!t]
        \centering
        \setlength{\tabcolsep}{1pt}
        \newcommand{\imgwidth}{0.18\linewidth}
        \begin{tabular}{cccccc}

        \multirow{-8}{*}{(a)} &
        \includegraphics[width=\imgwidth]{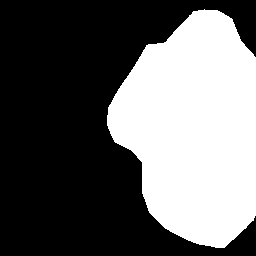} &
        \includegraphics[width=\imgwidth]{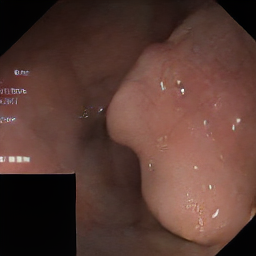} &
        \includegraphics[width=\imgwidth]{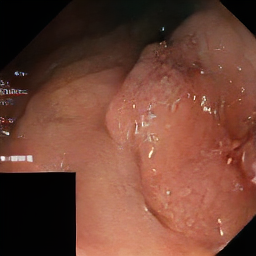} &
        \includegraphics[width=\imgwidth]{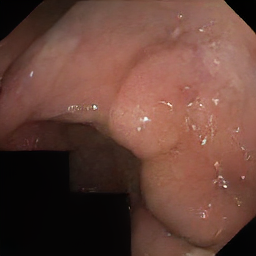} &
        \includegraphics[width=\imgwidth]{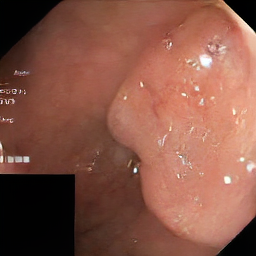} 
       
        \\

          \multirow{-8}{*}{(b)} &
        \includegraphics[width=\imgwidth]{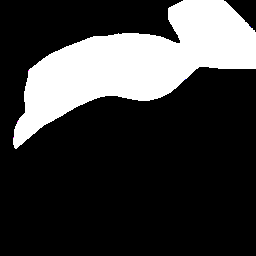} &
        \includegraphics[width=\imgwidth]{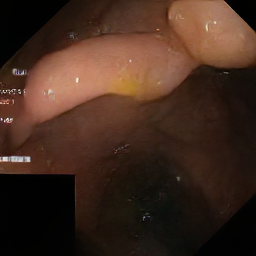} &
        \includegraphics[width=\imgwidth]{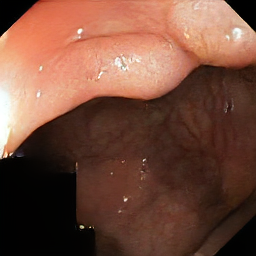} &
        \includegraphics[width=\imgwidth]{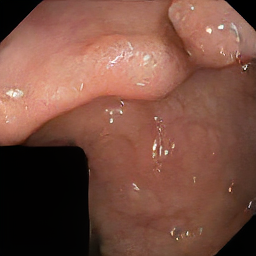} &
        \includegraphics[width=\imgwidth]{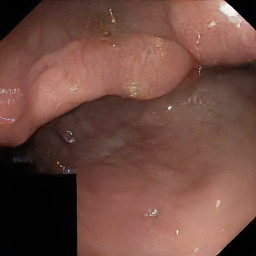} 
       
        \\

          \multirow{-8}{*}{(c)} &
        \includegraphics[width=\imgwidth]{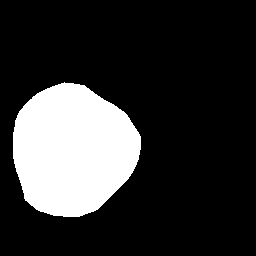} &
        \includegraphics[width=\imgwidth]{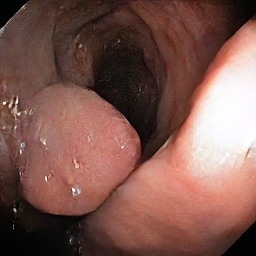} &
        \includegraphics[width=\imgwidth]{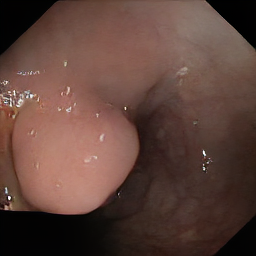} &
        \includegraphics[width=\imgwidth]{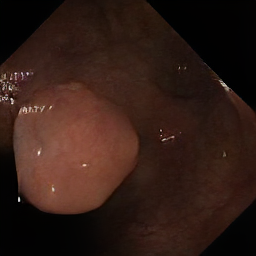} &
        \includegraphics[width=\imgwidth]{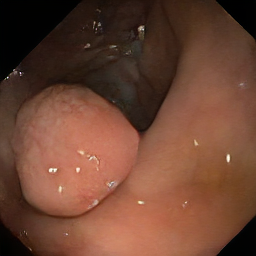} 
       
        \\

          \multirow{-8}{*}{(d)} &
        \includegraphics[width=\imgwidth]{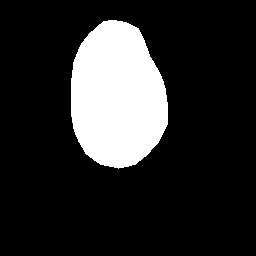} &
        \includegraphics[width=\imgwidth]{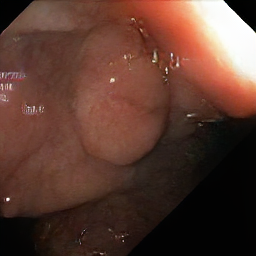} &
        \includegraphics[width=\imgwidth]{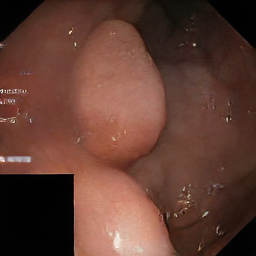} &
        \includegraphics[width=\imgwidth]{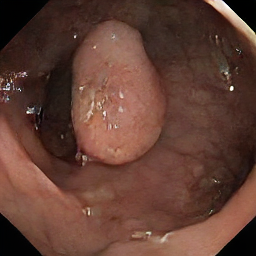} &
        \includegraphics[width=\imgwidth]{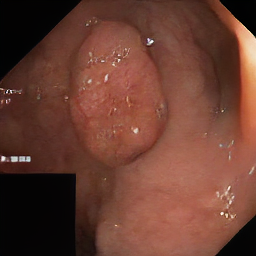}

        \end{tabular}

   \caption{Generated synthetic polyp images from our conditional probabilistic diffusion model. The first column shows input conditions to the latent diffusion model. All other columns show the corresponding stochastic polyps generations with different input noises.  }
    \label{fig:generated_polyps_multiple}
    
\end{figure*}

Therefore, we selected the model from earlier stages that achieved lowest \gls{fid}. We conditioned the model on one mask and generated multiple samples to see if the model generalizes well, results of this experiment can be seen in Figure \ref{fig:generated_polyps_multiple}.

\subsection{Segmentation experiments and results}
We used a learning rate of $0.0001$ with the Adam optimizer~\cite{kingma2014adam} to train the three segmenation models, UNet++, \gls{fpn} and DeepLabv3+. DiceLoss~\cite{sudre2017generalised} was used in the training process as the loss function to update the weights. The encoder model of \emph{resnet34}  was input as the encoder network for all three models (for more details of these encoder networks, please refer to the documentation~\cite{Iakubovskii:2019}). Micro metrics and micro-imagewise metrics (as discussed in the Pytorch segmentation library) were calculated from the best checkpoints and the test dataset after training $50$ epochs for all the models. The calculated micro metrics are tabulated in Table~\ref{tbl:micro_values}, and micro-imagewise values are tabulated in Table~\ref{tbl:micro_imagewise_values}.

\begin{table*}[]
\caption{Micro metrics calculated on the test dataset (300 real images and masks). The best value of each column is highlighted using \underline{underlined} text. R = Real, Syn = Synthetic, Acc: = Accuracy and Prec: = Precision. }
\begin{tabular}{ll|rrrr|rrrr|rrrr}
\toprule
 & Model: & \multicolumn{4}{c}{Unet++ (26.1M)} & \multicolumn{4}{c}{FPN (23.2M)} & \multicolumn{4}{c}{DeepLabv3plus (22.4M)} \\
 \midrule
\multicolumn{1}{c}{\# R} & \multicolumn{1}{c}{\# Syn} & \multicolumn{1}{c}{IOU} & \multicolumn{1}{c}{F1} & \multicolumn{1}{c}{Acc:} & \multicolumn{1}{c}{Prec:} & \multicolumn{1}{c}{IOU} & \multicolumn{1}{c}{F1} & \multicolumn{1}{c}{Acc:} & \multicolumn{1}{c}{Prec:} & \multicolumn{1}{c}{IOU} & \multicolumn{1}{c}{F1} & \multicolumn{1}{c}{Acc:} & \multicolumn{1}{c}{Prec:} \\
\midrule
700 & 0 & 0.7471 & 0.8552 & \underline{0.9509} & 0.8535 & \underline{0.7663} & \underline{0.8677} & \underline{0.9571} & 0.8623 & 0.7457 & 0.8543 & 0.9492 & 0.8699 \\
0 & 1000 & 0.6852 & 0.8132 & 0.9375 & 0.8009 & 0.6784 & 0.8084 & 0.9276 & 0.7685 & 0.6580 & 0.7938 & 0.9301 & 0.7658 \\
700 & 1000 & 0.7151 & 0.8339 & 0.9421 & 0.8235 & 0.7300 & 0.8439 & 0.9481 & 0.8323 & 0.7401 & 0.8506 & 0.9492 & 0.8252 \\
100 & 0 & 0.6970 & 0.8215 & 0.9400 & 0.7912 & 0.6840 & 0.8123 & 0.9371 & 0.8209 & 0.6983 & 0.8224 & 0.9404 & 0.8304 \\
100 & 100 & 0.6937 & 0.8192 & 0.9382 & 0.7692 & 0.7304 & 0.8442 & 0.9501 & 0.8509 & 0.7200 & 0.8372 & 0.9466 & 0.8305 \\
100 & 200 & 0.7066 & 0.8281 & 0.9418 & \underline{0.8804} & 0.7382 & 0.8494 & 0.9466 & \underline{0.8763} & 0.7040 & 0.8263 & 0.9429 & 0.8383 \\
100 & 300 & 0.7309 & 0.8445 & 0.9488 & 0.8536 & 0.7269 & 0.8419 & 0.9459 & 0.8219 & \underline{0.7556} & \underline{0.8608} & \underline{0.9500} & 0.8521 \\
100 & 400 & 0.6830 & 0.8116 & 0.9386 & 0.8333 & 0.7304 & 0.8442 & 0.9459 & 0.8375 & 0.7298 & 0.8438 & 0.9450 & 0.8342 \\
100 & 500 & 0.6815 & 0.8106 & 0.9366 & 0.8152 & 0.7244 & 0.8402 & 0.9421 & 0.8209 & 0.7212 & 0.8380 & 0.9454 & 0.8427 \\
100 & 600 & 0.7083 & 0.8292 & 0.9432 & 0.8287 & 0.7284 & 0.8429 & 0.9491 & 0.8668 & 0.7037 & 0.8261 & 0.9392 & 0.8405 \\
100 & 700 & 0.7195 & 0.8369 & 0.9460 & 0.8420 & 0.7436 & 0.8530 & 0.9498 & 0.8107 & 0.7083 & 0.8292 & 0.9457 & 0.8347 \\
100 & 800 & 0.6752 & 0.8061 & 0.9402 & 0.8495 & 0.7387 & 0.8497 & 0.9462 & 0.8278 & 0.7338 & 0.8465 & 0.9476 & \underline{0.8770} \\
100 & 900 & 0.7069 & 0.8283 & 0.9441 & 0.8319 & 0.7290 & 0.8432 & 0.9463 & 0.8234 & 0.7116 & 0.8315 & 0.9413 & 0.8171 \\
100 & 1000 & \underline{0.7513} & \underline{0.8580} & 0.9506 & 0.8468 & 0.7214 & 0.8382 & 0.9457 & 0.8126 & 0.7154 & 0.8341 & 0.9401 & 0.8337 \\
\bottomrule
\end{tabular}
\label{tbl:micro_values}
\end{table*}

\begin{table*}[]
\caption{Micro-imagewise metrics calculated on the test dataset (300 real images and masks). These metrics take into account class imbalance. The best value of each column is highlighted using \underline{underlined} text. R = Real, Syn = Synthetic, Acc: = Accuracy and Prec: = Precision. }
\begin{tabular}{ll|rrrr|rrrr|rrrr}
\toprule
 & Model: & \multicolumn{4}{c}{Unet++ (26.1M)} & \multicolumn{4}{c}{FPN (23.2M)} & \multicolumn{4}{c}{DeepLabv3plus (22.4M)} \\
 \midrule
\# R & \# Syn & \multicolumn{1}{l}{IOU} & \multicolumn{1}{l}{F1} & \multicolumn{1}{l}{Acc:} & \multicolumn{1}{l}{Prec:} & \multicolumn{1}{l}{IOU} & \multicolumn{1}{l}{F1} & \multicolumn{1}{l}{Acc:} & \multicolumn{1}{l}{Prec:} & \multicolumn{1}{l}{IOU} & \multicolumn{1}{l}{F1} & \multicolumn{1}{l}{Acc:} & \multicolumn{1}{l}{Prec:} \\
\midrule
700 & 0 & \underline{0.7551} & \underline{0.8222} & \underline{0.9509} & 0.8742 & \underline{0.7681} & \underline{0.8429} & \underline{0.9571} & 0.8761 & 0.7528 & 0.8317 & 0.9492 & 0.8678 \\
0 & 1000 & 0.7232 & 0.8013 & 0.9375 & 0.8128 & 0.6977 & 0.7903 & 0.9276 & 0.7757 & 0.7018 & 0.7896 & 0.9301 & 0.8062 \\
700 & 1000 & 0.7442 & 0.8185 & 0.9421 & 0.8517 & 0.7371 & 0.8128 & 0.9481 & 0.8504 & \underline{0.7751} & \underline{0.8465} & 0.9492 & 0.8628 \\
100 & 0 & 0.7136 & 0.8005 & 0.9400 & 0.7985 & 0.6587 & 0.7613 & 0.9371 & 0.8039 & 0.7116 & 0.8040 & 0.9404 & 0.8506 \\
100 & 100 & 0.7146 & 0.7976 & 0.9382 & 0.7931 & 0.7357 & 0.8212 & 0.9501 & 0.8483 & 0.7183 & 0.8048 & 0.9466 & 0.8420 \\
100 & 200 & 0.7433 & 0.8193 & 0.9418 & \underline{0.8768} & 0.7000 & 0.7842 & 0.9466 & \underline{0.8856} & 0.7197 & 0.8031 & 0.9429 & 0.8554 \\
100 & 300 & 0.7392 & 0.8168 & 0.9488 & 0.8570 & 0.7302 & 0.8126 & 0.9459 & 0.8357 & 0.7337 & 0.8097 & \underline{0.9500} & 0.8503 \\
100 & 400 & 0.7097 & 0.7867 & 0.9386 & 0.8444 & 0.7512 & 0.8268 & 0.9459 & 0.8683 & 0.7366 & 0.8153 & 0.9450 & 0.8604 \\
100 & 500 & 0.7088 & 0.7874 & 0.9366 & 0.8551 & 0.7376 & 0.8200 & 0.9421 & 0.8410 & 0.7264 & 0.8085 & 0.9454 & 0.8369 \\
100 & 600 & 0.7238 & 0.7987 & 0.9432 & 0.8584 & 0.7348 & 0.8147 & 0.9491 & 0.8639 & 0.7054 & 0.7944 & 0.9392 & 0.8287 \\
100 & 700 & 0.7230 & 0.7988 & 0.9460 & 0.8471 & 0.7319 & 0.8147 & 0.9498 & 0.8010 & 0.7317 & 0.8135 & 0.9457 & 0.8502 \\
100 & 800 & 0.7081 & 0.7884 & 0.9402 & 0.8692 & 0.7502 & 0.8274 & 0.9462 & 0.8622 & 0.7548 & 0.8322 & 0.9476 & \underline{0.8725} \\
100 & 900 & 0.7244 & 0.8025 & 0.9441 & 0.8329 & 0.7441 & 0.8242 & 0.9463 & 0.8534 & 0.7314 & 0.8115 & 0.9413 & 0.8570 \\
100 & 1000 & 0.7385 & 0.8145 & 0.9506 & 0.8495 & 0.7302 & 0.8086 & 0.9457 & 0.8386 & 0.7343 & 0.8125 & 0.9401 & 0.8574 \\
\bottomrule
\end{tabular}
\label{tbl:micro_imagewise_values}
\end{table*}

\begin{figure*}[!t]
        \centering
        \setlength{\tabcolsep}{1pt}
        \newcommand{\imgwidth}{0.12\linewidth}
        \begin{tabular}{cccccccc}

        Image & GT & UNet++(base) & UNet++(*) & FPN (base) & FPN(*) & DLab (base) & DLab (*)

        \\

        \includegraphics[width=\imgwidth]{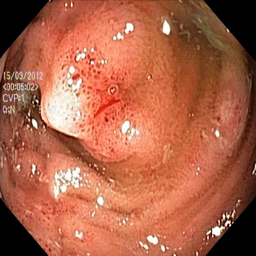} &
        \includegraphics[width=\imgwidth]{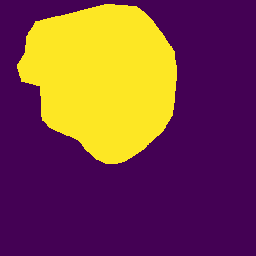} &
        \includegraphics[width=\imgwidth]{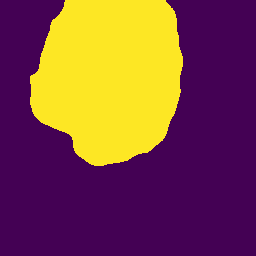} &
        \includegraphics[width=\imgwidth]{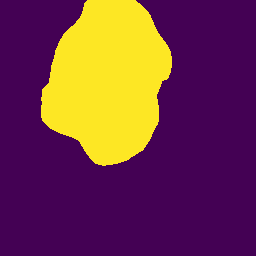} &
        \includegraphics[width=\imgwidth]{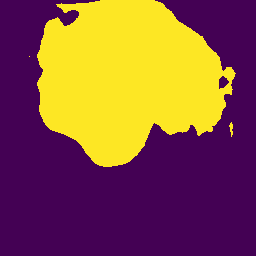} &
        \includegraphics[width=\imgwidth]{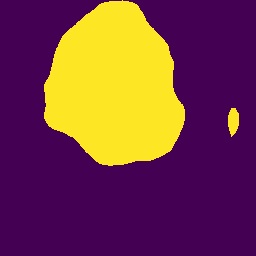} &
        \includegraphics[width=\imgwidth]{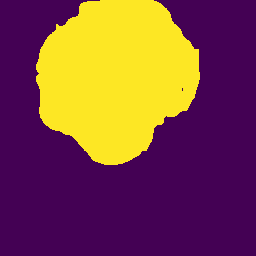} &
        \includegraphics[width=\imgwidth]{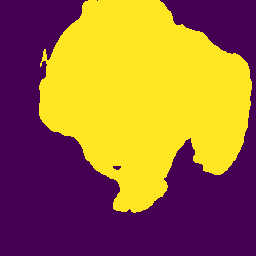} 
       
        \\

        \includegraphics[width=\imgwidth]{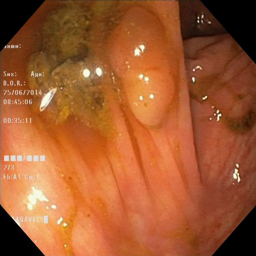} &
        \includegraphics[width=\imgwidth]{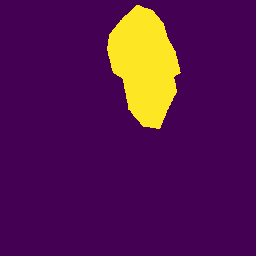} &
        \includegraphics[width=\imgwidth]{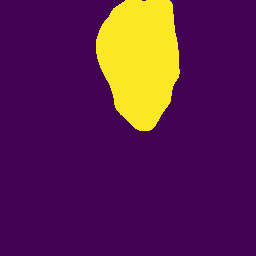} &
        \includegraphics[width=\imgwidth]{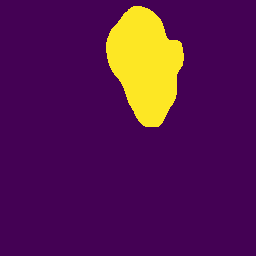} &
        \includegraphics[width=\imgwidth]{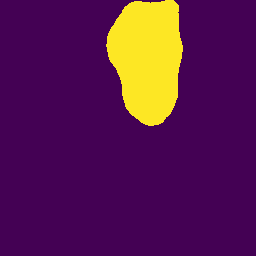} &
        \includegraphics[width=\imgwidth]{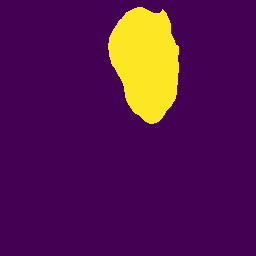} &
        \includegraphics[width=\imgwidth]{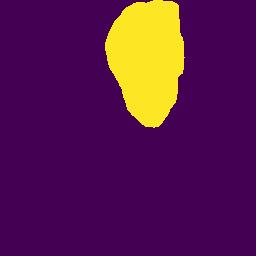} &
        \includegraphics[width=\imgwidth]{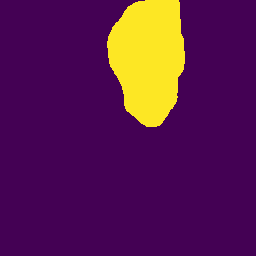} 
       
        \\

        \includegraphics[width=\imgwidth]{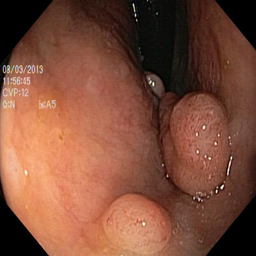} &
        \includegraphics[width=\imgwidth]{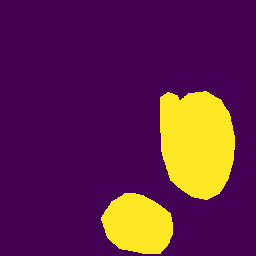} &
        \includegraphics[width=\imgwidth]{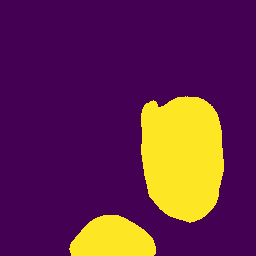} &
        \includegraphics[width=\imgwidth]{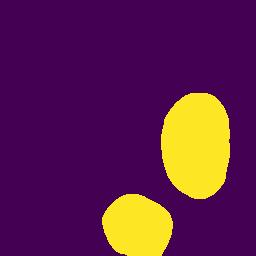} &
        \includegraphics[width=\imgwidth]{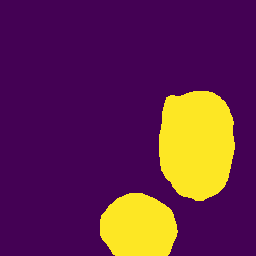} &
        \includegraphics[width=\imgwidth]{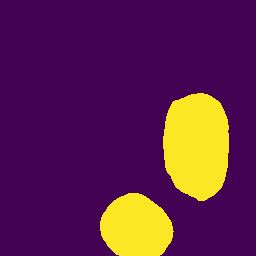} &
        \includegraphics[width=\imgwidth]{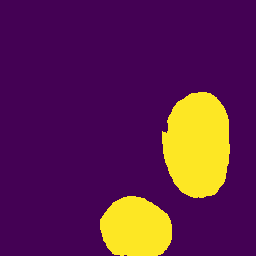} &
        \includegraphics[width=\imgwidth]{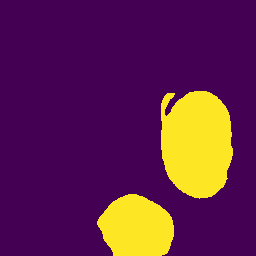} 
       
        \\

        \includegraphics[width=\imgwidth]{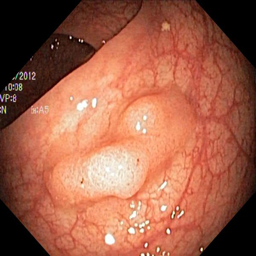} &
        \includegraphics[width=\imgwidth]{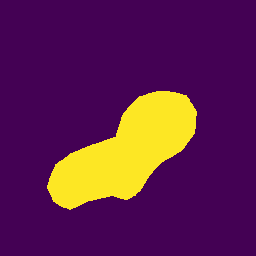} &
        \includegraphics[width=\imgwidth]{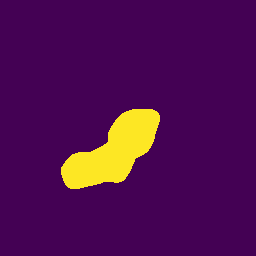} &
        \includegraphics[width=\imgwidth]{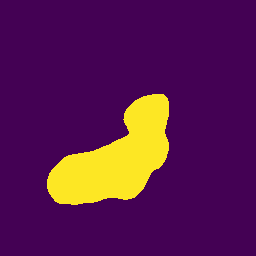} &
        \includegraphics[width=\imgwidth]{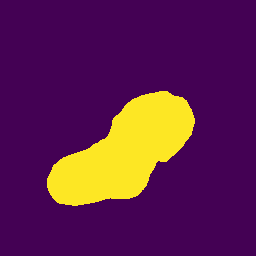} &
        \includegraphics[width=\imgwidth]{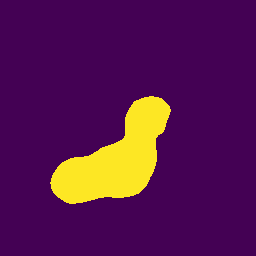} &
        \includegraphics[width=\imgwidth]{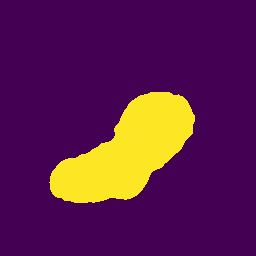} &
        \includegraphics[width=\imgwidth]{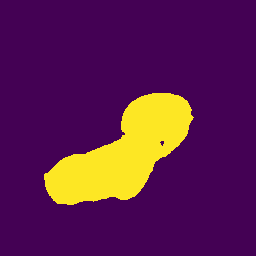}

        \end{tabular}

   \caption{Predicted masks from different segmentation models. The baseline predictions are from the model trained with only real data. Unet++(*) is selected based on using the high \gls{iou} value in Table \ref{tbl:micro_values}. FPN(*) is selected using the highest Precision in Tables~\ref{tbl:micro_values} and ~\ref{tbl:micro_imagewise_values}. DeepLabv3(*)[Dlab(*)] is selected using high \gls{iou} values in Table \ref{tbl:micro_imagewise_values}. }
    \label{fig:segmentation_predictions}
    
\end{figure*}

According to the results in Tables~\ref{tbl:micro_values} and \ref{tbl:micro_imagewise_values}, it is clear that adding synthetic data can improve the results of segmentation models. However, it is not always true because some models like \gls{fpn} and UNet++ show the best \gls{iou}, F1, and accuracy when the training data consists of only the real data. In contrast, DeepLabv3 shows the best performance when some synthetic data is included in the training data. Overall, the best micro-imagewise \gls{iou} of $0.7751$ is achieved from DeepLabv3+ when the training data contains the maximum number of images from both the real data and the synthetic data. Therefore, it is a clear evidence that synthetic data has a direct influence on the final performance of segmentation models. Moreover, we noticed that precision is always better when the synthetic data is in the training data than using only the real data. This implies that synthesized \gls{tp} can improve the \gls{tp} predictions which is more important in the medical domain. More visual comparisons are presented in Figure~\ref{fig:segmentation_predictions}. This figure compares the model predictions from the three segmentation models between baseline performance and improved performance (marked using * in Figure~\ref{fig:segmentation_predictions}) using synthetic data. The improved versions of the models were selected using the metrics of Tables~\ref{tbl:micro_values} and \ref{tbl:micro_imagewise_values}.  

Another interesting finding of these segmentation experiments is that we get the best values of precision, accuracy, F1 and \gls{iou} when we use a smaller number of real images and synthetic images. For example, Unet++ and \gls{fpn} shows best precision values (micro and micro-imagewise)  when the training data consist of $100$ real samples and $200$ synthetic samples . However, this implies that there is a direct correlation between synthetic data, models, and the number of parameters. Therefore, researchers should not conclude performance gain or degrade of using synthetic data to train segmentation models just by evaluating a single model. 

\section{Conclusion and future works}

In this study, we used a probabilistic diffusion model-based method to generate synthetic polyp images conditioned on synthetic polyp masks. Our visual and quantitative comparisons show that the generated synthetic data are unique and realistic and not a simple copy of the training data used for the training.  Our further analysis of using synthetic data to train polyp segmentation models shows that synthetic data can be used to improve the performance of segmentation models while these improvements are correlated with model architectures. In this regard, we can clearly conclude that synthetic data help to improve the performance of segmentation models. However, deep evaluations should be performed with multiple model architecture to see the real gain of using synthetic data.

In future studies, we will perform more segmentation experiments to get a complete result set for Tables~\ref{tbl:micro_values} and~\ref{tbl:micro_imagewise_values}, for example, increasing the synthetic training data gradually with the full real dataset. Moreover, we will generate more synthetic data using our model to train the segmentation models with large synthetic datasets to evaluate the effect of using synthetic data deeply. Generating multiple images conditioned on the same input to train the segmentation models are an another limitation of the presented segmentation experiments. Furthermore, the quality of generated images can be improved using the style-transfer technique~\cite{gatys2016image} as used in the SinGAN-Seg study~\cite{thambawita2022singan}. Cross-dataset evaluations should be performed to measure the effect of using synthetic data to train segmentation models to improve robustness and generalizability.   

\begin{acks}
The research presented in this paper has benefited from the Experimental Infrastructure for Exploration of Exascale Computing (eX3), which is financially supported by the Research Council of Norway under contract 270053.
\end{acks}

\bibliographystyle{ACM-Reference-Format}
\bibliography{references}


\begin{thebibliography}{44}


\ifx \showCODEN    \undefined \def \showCODEN     #1{\unskip}     \fi
\ifx \showDOI      \undefined \def \showDOI       #1{#1}\fi
\ifx \showISBNx    \undefined \def \showISBNx     #1{\unskip}     \fi
\ifx \showISBNxiii \undefined \def \showISBNxiii  #1{\unskip}     \fi
\ifx \showISSN     \undefined \def \showISSN      #1{\unskip}     \fi
\ifx \showLCCN     \undefined \def \showLCCN      #1{\unskip}     \fi
\ifx \shownote     \undefined \def \shownote      #1{#1}          \fi
\ifx \showarticletitle \undefined \def \showarticletitle #1{#1}   \fi
\ifx \showURL      \undefined \def \showURL       {\relax}        \fi
\providecommand\bibfield[2]{#2}
\providecommand\bibinfo[2]{#2}
\providecommand\natexlab[1]{#1}
\providecommand\showeprint[2][]{arXiv:#2}

\bibitem[Adjei et~al\mbox{.}(2022)]%
        {adjei2022examining}
\bibfield{author}{\bibinfo{person}{Prince~Ebenezer Adjei},
  \bibinfo{person}{Zenebe~Markos Lonseko}, \bibinfo{person}{Wenju Du},
  \bibinfo{person}{Han Zhang}, {and} \bibinfo{person}{Nini Rao}.}
  \bibinfo{year}{2022}\natexlab{}.
\newblock \showarticletitle{Examining the effect of synthetic data augmentation
  in polyp detection and segmentation}.
\newblock \bibinfo{journal}{\emph{International Journal of Computer Assisted
  Radiology and Surgery}} \bibinfo{volume}{17}, \bibinfo{number}{7}
  (\bibinfo{year}{2022}), \bibinfo{pages}{1289--1302}.
\newblock


\bibitem[Alqahtani et~al\mbox{.}(2021)]%
        {alqahtani2021applications}
\bibfield{author}{\bibinfo{person}{Hamed Alqahtani}, \bibinfo{person}{Manolya
  Kavakli-Thorne}, {and} \bibinfo{person}{Gulshan Kumar}.}
  \bibinfo{year}{2021}\natexlab{}.
\newblock \showarticletitle{Applications of generative adversarial networks
  (gans): An updated review}.
\newblock \bibinfo{journal}{\emph{Archives of Computational Methods in
  Engineering}}  \bibinfo{volume}{28} (\bibinfo{year}{2021}),
  \bibinfo{pages}{525--552}.
\newblock


\bibitem[Bernal et~al\mbox{.}(2015)]%
        {bernal2015wm}
\bibfield{author}{\bibinfo{person}{Jorge Bernal}, \bibinfo{person}{F~Javier
  S{\'a}nchez}, \bibinfo{person}{Gloria Fern{\'a}ndez-Esparrach},
  \bibinfo{person}{Debora Gil}, \bibinfo{person}{Cristina Rodr{\'\i}guez},
  {and} \bibinfo{person}{Fernando Vilari{\~n}o}.}
  \bibinfo{year}{2015}\natexlab{}.
\newblock \showarticletitle{WM-DOVA maps for accurate polyp highlighting in
  colonoscopy: Validation vs. saliency maps from physicians}.
\newblock \bibinfo{journal}{\emph{Computerized medical imaging and graphics}}
  \bibinfo{volume}{43} (\bibinfo{year}{2015}), \bibinfo{pages}{99--111}.
\newblock


\bibitem[Borgli et~al\mbox{.}(2020)]%
        {borgli2020hyperkvasir}
\bibfield{author}{\bibinfo{person}{Hanna Borgli}, \bibinfo{person}{Vajira
  Thambawita}, \bibinfo{person}{Pia~H Smedsrud}, \bibinfo{person}{Steven
  Hicks}, \bibinfo{person}{Debesh Jha}, \bibinfo{person}{Sigrun~L Eskeland},
  \bibinfo{person}{Kristin~Ranheim Randel}, \bibinfo{person}{Konstantin
  Pogorelov}, \bibinfo{person}{Mathias Lux}, \bibinfo{person}{Duc Tien~Dang
  Nguyen}, {et~al\mbox{.}}} \bibinfo{year}{2020}\natexlab{}.
\newblock \showarticletitle{HyperKvasir, a comprehensive multi-class image and
  video dataset for gastrointestinal endoscopy}.
\newblock \bibinfo{journal}{\emph{Scientific data}} \bibinfo{volume}{7},
  \bibinfo{number}{1} (\bibinfo{year}{2020}), \bibinfo{pages}{283}.
\newblock


\bibitem[Chen et~al\mbox{.}(2017)]%
        {chen2017rethinking}
\bibfield{author}{\bibinfo{person}{Liang-Chieh Chen}, \bibinfo{person}{George
  Papandreou}, \bibinfo{person}{Florian Schroff}, {and}
  \bibinfo{person}{Hartwig Adam}.} \bibinfo{year}{2017}\natexlab{}.
\newblock \showarticletitle{Rethinking atrous convolution for semantic image
  segmentation}.
\newblock \bibinfo{journal}{\emph{arXiv preprint arXiv:1706.05587}}
  (\bibinfo{year}{2017}).
\newblock


\bibitem[Chen et~al\mbox{.}(2021)]%
        {chen2021synthetic}
\bibfield{author}{\bibinfo{person}{Richard~J Chen}, \bibinfo{person}{Ming~Y
  Lu}, \bibinfo{person}{Tiffany~Y Chen}, \bibinfo{person}{Drew~FK Williamson},
  {and} \bibinfo{person}{Faisal Mahmood}.} \bibinfo{year}{2021}\natexlab{}.
\newblock \showarticletitle{Synthetic data in machine learning for medicine and
  healthcare}.
\newblock \bibinfo{journal}{\emph{Nature Biomedical Engineering}}
  \bibinfo{volume}{5}, \bibinfo{number}{6} (\bibinfo{year}{2021}),
  \bibinfo{pages}{493--497}.
\newblock


\bibitem[Creswell et~al\mbox{.}(2018)]%
        {creswell2018generative}
\bibfield{author}{\bibinfo{person}{Antonia Creswell}, \bibinfo{person}{Tom
  White}, \bibinfo{person}{Vincent Dumoulin}, \bibinfo{person}{Kai
  Arulkumaran}, \bibinfo{person}{Biswa Sengupta}, {and} \bibinfo{person}{Anil~A
  Bharath}.} \bibinfo{year}{2018}\natexlab{}.
\newblock \showarticletitle{Generative adversarial networks: An overview}.
\newblock \bibinfo{journal}{\emph{IEEE signal processing magazine}}
  \bibinfo{volume}{35}, \bibinfo{number}{1} (\bibinfo{year}{2018}),
  \bibinfo{pages}{53--65}.
\newblock


\bibitem[Dhariwal and Nichol(2021)]%
        {dhariwal2021diffusion}
\bibfield{author}{\bibinfo{person}{Prafulla Dhariwal} {and}
  \bibinfo{person}{Alexander Nichol}.} \bibinfo{year}{2021}\natexlab{}.
\newblock \showarticletitle{Diffusion models beat gans on image synthesis}.
\newblock \bibinfo{journal}{\emph{Advances in Neural Information Processing
  Systems}}  \bibinfo{volume}{34} (\bibinfo{year}{2021}),
  \bibinfo{pages}{8780--8794}.
\newblock


\bibitem[Elharrouss et~al\mbox{.}(2020)]%
        {elharrouss2020image}
\bibfield{author}{\bibinfo{person}{Omar Elharrouss}, \bibinfo{person}{Noor
  Almaadeed}, \bibinfo{person}{Somaya Al-Maadeed}, {and}
  \bibinfo{person}{Younes Akbari}.} \bibinfo{year}{2020}\natexlab{}.
\newblock \showarticletitle{Image inpainting: A review}.
\newblock \bibinfo{journal}{\emph{Neural Processing Letters}}
  \bibinfo{volume}{51} (\bibinfo{year}{2020}), \bibinfo{pages}{2007--2028}.
\newblock


\bibitem[Fagereng et~al\mbox{.}(2022)]%
        {fagereng2022polypconnect}
\bibfield{author}{\bibinfo{person}{Jan~Andre Fagereng}, \bibinfo{person}{Vajira
  Thambawita}, \bibinfo{person}{Andrea~M Stor{\aa}s},
  \bibinfo{person}{Sravanthi Parasa}, \bibinfo{person}{Thomas de Lange},
  \bibinfo{person}{P{\aa}l Halvorsen}, {and} \bibinfo{person}{Michael~A
  Riegler}.} \bibinfo{year}{2022}\natexlab{}.
\newblock \showarticletitle{PolypConnect: Image inpainting for generating
  realistic gastrointestinal tract images with polyps}. In
  \bibinfo{booktitle}{\emph{2022 IEEE 35th International Symposium on
  Computer-Based Medical Systems (CBMS)}}. IEEE, \bibinfo{pages}{66--71}.
\newblock


\bibitem[Gatys et~al\mbox{.}(2016)]%
        {gatys2016image}
\bibfield{author}{\bibinfo{person}{Leon~A Gatys}, \bibinfo{person}{Alexander~S
  Ecker}, {and} \bibinfo{person}{Matthias Bethge}.}
  \bibinfo{year}{2016}\natexlab{}.
\newblock \showarticletitle{Image style transfer using convolutional neural
  networks}. In \bibinfo{booktitle}{\emph{Proceedings of the IEEE conference on
  computer vision and pattern recognition}}. \bibinfo{pages}{2414--2423}.
\newblock


\bibitem[Heusel et~al\mbox{.}(2017)]%
        {heusel2017gans}
\bibfield{author}{\bibinfo{person}{Martin Heusel}, \bibinfo{person}{Hubert
  Ramsauer}, \bibinfo{person}{Thomas Unterthiner}, \bibinfo{person}{Bernhard
  Nessler}, {and} \bibinfo{person}{Sepp Hochreiter}.}
  \bibinfo{year}{2017}\natexlab{}.
\newblock \showarticletitle{Gans trained by a two time-scale update rule
  converge to a local nash equilibrium}.
\newblock \bibinfo{journal}{\emph{Advances in neural information processing
  systems}}  \bibinfo{volume}{30} (\bibinfo{year}{2017}).
\newblock


\bibitem[Ho et~al\mbox{.}(2020)]%
        {ho2020denoising}
\bibfield{author}{\bibinfo{person}{Jonathan Ho}, \bibinfo{person}{Ajay Jain},
  {and} \bibinfo{person}{Pieter Abbeel}.} \bibinfo{year}{2020}\natexlab{}.
\newblock \showarticletitle{Denoising diffusion probabilistic models}.
\newblock \bibinfo{journal}{\emph{Advances in Neural Information Processing
  Systems}}  \bibinfo{volume}{33} (\bibinfo{year}{2020}),
  \bibinfo{pages}{6840--6851}.
\newblock


\bibitem[Iakubovskii(2019)]%
        {Iakubovskii:2019}
\bibfield{author}{\bibinfo{person}{Pavel Iakubovskii}.}
  \bibinfo{year}{2019}\natexlab{}.
\newblock \bibinfo{title}{Segmentation Models Pytorch}.
\newblock
  \bibinfo{howpublished}{\url{https://github.com/qubvel/segmentation_models.pytorch}}.
\newblock


\bibitem[Iddan et~al\mbox{.}(2000)]%
        {iddan2000wireless}
\bibfield{author}{\bibinfo{person}{Gavriel Iddan}, \bibinfo{person}{Gavriel
  Meron}, \bibinfo{person}{Arkady Glukhovsky}, {and} \bibinfo{person}{Paul
  Swain}.} \bibinfo{year}{2000}\natexlab{}.
\newblock \showarticletitle{Wireless capsule endoscopy}.
\newblock \bibinfo{journal}{\emph{Nature}} \bibinfo{volume}{405},
  \bibinfo{number}{6785} (\bibinfo{year}{2000}), \bibinfo{pages}{417--417}.
\newblock


\bibitem[Jha et~al\mbox{.}(2020)]%
        {jha2020kvasir}
\bibfield{author}{\bibinfo{person}{Debesh Jha}, \bibinfo{person}{Pia~H
  Smedsrud}, \bibinfo{person}{Michael~A Riegler}, \bibinfo{person}{P{\aa}l
  Halvorsen}, \bibinfo{person}{Thomas de Lange}, \bibinfo{person}{Dag
  Johansen}, {and} \bibinfo{person}{H{\aa}vard~D Johansen}.}
  \bibinfo{year}{2020}\natexlab{}.
\newblock \showarticletitle{Kvasir-seg: A segmented polyp dataset}. In
  \bibinfo{booktitle}{\emph{MultiMedia Modeling: 26th International Conference,
  MMM 2020, Daejeon, South Korea, January 5--8, 2020, Proceedings, Part II
  26}}. Springer, \bibinfo{pages}{451--462}.
\newblock


\bibitem[Jordon et~al\mbox{.}(2022)]%
        {jordon2022synthetic}
\bibfield{author}{\bibinfo{person}{James Jordon}, \bibinfo{person}{Lukasz
  Szpruch}, \bibinfo{person}{Florimond Houssiau}, \bibinfo{person}{Mirko
  Bottarelli}, \bibinfo{person}{Giovanni Cherubin}, \bibinfo{person}{Carsten
  Maple}, \bibinfo{person}{Samuel~N Cohen}, {and} \bibinfo{person}{Adrian
  Weller}.} \bibinfo{year}{2022}\natexlab{}.
\newblock \showarticletitle{Synthetic Data--what, why and how?}
\newblock \bibinfo{journal}{\emph{arXiv preprint arXiv:2205.03257}}
  (\bibinfo{year}{2022}).
\newblock


\bibitem[Kaminski et~al\mbox{.}(2010)]%
        {kaminski2010quality}
\bibfield{author}{\bibinfo{person}{Michal~F Kaminski},
  \bibinfo{person}{Jaroslaw Regula}, \bibinfo{person}{Ewa Kraszewska},
  \bibinfo{person}{Marcin Polkowski}, \bibinfo{person}{Urszula Wojciechowska},
  \bibinfo{person}{Joanna Didkowska}, \bibinfo{person}{Maria Zwierko},
  \bibinfo{person}{Maciej Rupinski}, \bibinfo{person}{Marek~P Nowacki}, {and}
  \bibinfo{person}{Eugeniusz Butruk}.} \bibinfo{year}{2010}\natexlab{}.
\newblock \showarticletitle{Quality indicators for colonoscopy and the risk of
  interval cancer}.
\newblock \bibinfo{journal}{\emph{New England journal of medicine}}
  \bibinfo{volume}{362}, \bibinfo{number}{19} (\bibinfo{year}{2010}),
  \bibinfo{pages}{1795--1803}.
\newblock


\bibitem[Kingma and Ba(2014)]%
        {kingma2014adam}
\bibfield{author}{\bibinfo{person}{Diederik~P Kingma} {and}
  \bibinfo{person}{Jimmy Ba}.} \bibinfo{year}{2014}\natexlab{}.
\newblock \showarticletitle{Adam: A method for stochastic optimization}.
\newblock \bibinfo{journal}{\emph{arXiv preprint arXiv:1412.6980}}
  (\bibinfo{year}{2014}).
\newblock


\bibitem[Krizhevsky et~al\mbox{.}(2017)]%
        {krizhevsky2017imagenet}
\bibfield{author}{\bibinfo{person}{Alex Krizhevsky}, \bibinfo{person}{Ilya
  Sutskever}, {and} \bibinfo{person}{Geoffrey~E Hinton}.}
  \bibinfo{year}{2017}\natexlab{}.
\newblock \showarticletitle{Imagenet classification with deep convolutional
  neural networks}.
\newblock \bibinfo{journal}{\emph{Commun. ACM}} \bibinfo{volume}{60},
  \bibinfo{number}{6} (\bibinfo{year}{2017}), \bibinfo{pages}{84--90}.
\newblock


\bibitem[Le~Berre et~al\mbox{.}(2020)]%
        {le2020application}
\bibfield{author}{\bibinfo{person}{Catherine Le~Berre},
  \bibinfo{person}{William~J Sandborn}, \bibinfo{person}{Sabeur Aridhi},
  \bibinfo{person}{Marie-Dominique Devignes}, \bibinfo{person}{Laure Fournier},
  \bibinfo{person}{Malika Smail-Tabbone}, \bibinfo{person}{Silvio Danese},
  {and} \bibinfo{person}{Laurent Peyrin-Biroulet}.}
  \bibinfo{year}{2020}\natexlab{}.
\newblock \showarticletitle{Application of artificial intelligence to
  gastroenterology and hepatology}.
\newblock \bibinfo{journal}{\emph{Gastroenterology}} \bibinfo{volume}{158},
  \bibinfo{number}{1} (\bibinfo{year}{2020}), \bibinfo{pages}{76--94}.
\newblock


\bibitem[Lin et~al\mbox{.}(2017)]%
        {lin2017feature}
\bibfield{author}{\bibinfo{person}{Tsung-Yi Lin}, \bibinfo{person}{Piotr
  Doll{\'a}r}, \bibinfo{person}{Ross Girshick}, \bibinfo{person}{Kaiming He},
  \bibinfo{person}{Bharath Hariharan}, {and} \bibinfo{person}{Serge Belongie}.}
  \bibinfo{year}{2017}\natexlab{}.
\newblock \showarticletitle{Feature pyramid networks for object detection}. In
  \bibinfo{booktitle}{\emph{Proceedings of the IEEE conference on computer
  vision and pattern recognition}}. \bibinfo{pages}{2117--2125}.
\newblock


\bibitem[Lin et~al\mbox{.}(2014)]%
        {lin2014microsoft}
\bibfield{author}{\bibinfo{person}{Tsung-Yi Lin}, \bibinfo{person}{Michael
  Maire}, \bibinfo{person}{Serge Belongie}, \bibinfo{person}{James Hays},
  \bibinfo{person}{Pietro Perona}, \bibinfo{person}{Deva Ramanan},
  \bibinfo{person}{Piotr Doll{\'a}r}, {and} \bibinfo{person}{C~Lawrence
  Zitnick}.} \bibinfo{year}{2014}\natexlab{}.
\newblock \showarticletitle{Microsoft coco: Common objects in context}. In
  \bibinfo{booktitle}{\emph{Computer Vision--ECCV 2014: 13th European
  Conference, Zurich, Switzerland, September 6-12, 2014, Proceedings, Part V
  13}}. Springer, \bibinfo{pages}{740--755}.
\newblock


\bibitem[Min et~al\mbox{.}(2019)]%
        {min2019computer}
\bibfield{author}{\bibinfo{person}{Min Min}, \bibinfo{person}{Song Su},
  \bibinfo{person}{Wenrui He}, \bibinfo{person}{Yiliang Bi},
  \bibinfo{person}{Zhanyu Ma}, {and} \bibinfo{person}{Yan Liu}.}
  \bibinfo{year}{2019}\natexlab{}.
\newblock \showarticletitle{Computer-aided diagnosis of colorectal polyps using
  linked color imaging colonoscopy to predict histology}.
\newblock \bibinfo{journal}{\emph{Scientific reports}} \bibinfo{volume}{9},
  \bibinfo{number}{1} (\bibinfo{year}{2019}), \bibinfo{pages}{1--8}.
\newblock


\bibitem[Nichol and Dhariwal(2021)]%
        {nichol2021improved}
\bibfield{author}{\bibinfo{person}{Alexander~Quinn Nichol} {and}
  \bibinfo{person}{Prafulla Dhariwal}.} \bibinfo{year}{2021}\natexlab{}.
\newblock \showarticletitle{Improved denoising diffusion probabilistic models}.
  In \bibinfo{booktitle}{\emph{International Conference on Machine Learning}}.
  PMLR, \bibinfo{pages}{8162--8171}.
\newblock


\bibitem[Paszke et~al\mbox{.}(2019)]%
        {NEURIPS2019_9015}
\bibfield{author}{\bibinfo{person}{Adam Paszke}, \bibinfo{person}{Sam Gross},
  \bibinfo{person}{Francisco Massa}, \bibinfo{person}{Adam Lerer},
  \bibinfo{person}{James Bradbury}, \bibinfo{person}{Gregory Chanan},
  \bibinfo{person}{Trevor Killeen}, \bibinfo{person}{Zeming Lin},
  \bibinfo{person}{Natalia Gimelshein}, \bibinfo{person}{Luca Antiga},
  \bibinfo{person}{Alban Desmaison}, \bibinfo{person}{Andreas Kopf},
  \bibinfo{person}{Edward Yang}, \bibinfo{person}{Zachary DeVito},
  \bibinfo{person}{Martin Raison}, \bibinfo{person}{Alykhan Tejani},
  \bibinfo{person}{Sasank Chilamkurthy}, \bibinfo{person}{Benoit Steiner},
  \bibinfo{person}{Lu Fang}, \bibinfo{person}{Junjie Bai}, {and}
  \bibinfo{person}{Soumith Chintala}.} \bibinfo{year}{2019}\natexlab{}.
\newblock \showarticletitle{PyTorch: An Imperative Style, High-Performance Deep
  Learning Library}.
\newblock In \bibinfo{booktitle}{\emph{Advances in Neural Information
  Processing Systems 32}}. \bibinfo{publisher}{Curran Associates, Inc.},
  \bibinfo{pages}{8024--8035}.
\newblock
\urldef\tempurl%
\url{http://papers.neurips.cc/paper/9015-pytorch-an-imperative-style-high-performance-deep-learning-library.pdf}
\showURL{%
\tempurl}


\bibitem[Pogorelov et~al\mbox{.}(2017)]%
        {pogorelov2017kvasir}
\bibfield{author}{\bibinfo{person}{Konstantin Pogorelov},
  \bibinfo{person}{Kristin~Ranheim Randel}, \bibinfo{person}{Carsten Griwodz},
  \bibinfo{person}{Sigrun~Losada Eskeland}, \bibinfo{person}{Thomas de Lange},
  \bibinfo{person}{Dag Johansen}, \bibinfo{person}{Concetto Spampinato},
  \bibinfo{person}{Duc-Tien Dang-Nguyen}, \bibinfo{person}{Mathias Lux},
  \bibinfo{person}{Peter~Thelin Schmidt}, {et~al\mbox{.}}}
  \bibinfo{year}{2017}\natexlab{}.
\newblock \showarticletitle{Kvasir: A multi-class image dataset for computer
  aided gastrointestinal disease detection}. In
  \bibinfo{booktitle}{\emph{Proceedings of the 8th ACM on Multimedia Systems
  Conference}}. \bibinfo{pages}{164--169}.
\newblock


\bibitem[Qadir et~al\mbox{.}(2022)]%
        {qadir2022simple}
\bibfield{author}{\bibinfo{person}{Hemin~Ali Qadir}, \bibinfo{person}{Ilangko
  Balasingham}, {and} \bibinfo{person}{Younghak Shin}.}
  \bibinfo{year}{2022}\natexlab{}.
\newblock \showarticletitle{Simple U-net based synthetic polyp image
  generation: Polyp to negative and negative to polyp}.
\newblock \bibinfo{journal}{\emph{Biomedical Signal Processing and Control}}
  \bibinfo{volume}{74} (\bibinfo{year}{2022}), \bibinfo{pages}{103491}.
\newblock


\bibitem[Riegler et~al\mbox{.}(2016)]%
        {riegler2016eir}
\bibfield{author}{\bibinfo{person}{Michael Riegler},
  \bibinfo{person}{Konstantin Pogorelov}, \bibinfo{person}{P{\aa}l Halvorsen},
  \bibinfo{person}{Thomas de Lange}, \bibinfo{person}{Carsten Griwodz},
  \bibinfo{person}{Peter~Thelin Schmidt}, \bibinfo{person}{Sigrun~Losada
  Eskeland}, {and} \bibinfo{person}{Dag Johansen}.}
  \bibinfo{year}{2016}\natexlab{}.
\newblock \showarticletitle{Eir—efficient computer aided diagnosis framework
  for gastrointestinal endoscopies}. In \bibinfo{booktitle}{\emph{2016 14th
  International Workshop on Content-Based Multimedia Indexing (CBMI)}}. IEEE,
  \bibinfo{pages}{1--6}.
\newblock


\bibitem[Rombach et~al\mbox{.}(2021)]%
        {rombach2021highresolution}
\bibfield{author}{\bibinfo{person}{Robin Rombach}, \bibinfo{person}{Andreas
  Blattmann}, \bibinfo{person}{Dominik Lorenz}, \bibinfo{person}{Patrick
  Esser}, {and} \bibinfo{person}{Björn Ommer}.}
  \bibinfo{year}{2021}\natexlab{}.
\newblock \bibinfo{title}{High-Resolution Image Synthesis with Latent Diffusion
  Models}.
\newblock
\newblock
\showeprint[arxiv]{2112.10752}~[cs.CV]


\bibitem[Rombach et~al\mbox{.}(2022)]%
        {rombach2022high}
\bibfield{author}{\bibinfo{person}{Robin Rombach}, \bibinfo{person}{Andreas
  Blattmann}, \bibinfo{person}{Dominik Lorenz}, \bibinfo{person}{Patrick
  Esser}, {and} \bibinfo{person}{Bj{\"o}rn Ommer}.}
  \bibinfo{year}{2022}\natexlab{}.
\newblock \showarticletitle{High-resolution image synthesis with latent
  diffusion models}. In \bibinfo{booktitle}{\emph{Proceedings of the IEEE/CVF
  Conference on Computer Vision and Pattern Recognition}}.
  \bibinfo{pages}{10684--10695}.
\newblock


\bibitem[Schuhmann et~al\mbox{.}(2021)]%
        {schuhmann2021laion}
\bibfield{author}{\bibinfo{person}{Christoph Schuhmann},
  \bibinfo{person}{Richard Vencu}, \bibinfo{person}{Romain Beaumont},
  \bibinfo{person}{Robert Kaczmarczyk}, \bibinfo{person}{Clayton Mullis},
  \bibinfo{person}{Aarush Katta}, \bibinfo{person}{Theo Coombes},
  \bibinfo{person}{Jenia Jitsev}, {and} \bibinfo{person}{Aran Komatsuzaki}.}
  \bibinfo{year}{2021}\natexlab{}.
\newblock \showarticletitle{Laion-400m: Open dataset of clip-filtered 400
  million image-text pairs}.
\newblock \bibinfo{journal}{\emph{arXiv preprint arXiv:2111.02114}}
  (\bibinfo{year}{2021}).
\newblock


\bibitem[Silva et~al\mbox{.}(2014)]%
        {silva2014toward}
\bibfield{author}{\bibinfo{person}{Juan Silva}, \bibinfo{person}{Aymeric
  Histace}, \bibinfo{person}{Olivier Romain}, \bibinfo{person}{Xavier Dray},
  {and} \bibinfo{person}{Bertrand Granado}.} \bibinfo{year}{2014}\natexlab{}.
\newblock \showarticletitle{Toward embedded detection of polyps in wce images
  for early diagnosis of colorectal cancer}.
\newblock \bibinfo{journal}{\emph{International journal of computer assisted
  radiology and surgery}}  \bibinfo{volume}{9} (\bibinfo{year}{2014}),
  \bibinfo{pages}{283--293}.
\newblock


\bibitem[Sohl-Dickstein et~al\mbox{.}(2015)]%
        {sohl2015deep}
\bibfield{author}{\bibinfo{person}{Jascha Sohl-Dickstein},
  \bibinfo{person}{Eric Weiss}, \bibinfo{person}{Niru Maheswaranathan}, {and}
  \bibinfo{person}{Surya Ganguli}.} \bibinfo{year}{2015}\natexlab{}.
\newblock \showarticletitle{Deep unsupervised learning using nonequilibrium
  thermodynamics}. In \bibinfo{booktitle}{\emph{International Conference on
  Machine Learning}}. PMLR, \bibinfo{pages}{2256--2265}.
\newblock


\bibitem[Sudre et~al\mbox{.}(2017)]%
        {sudre2017generalised}
\bibfield{author}{\bibinfo{person}{Carole~H Sudre}, \bibinfo{person}{Wenqi Li},
  \bibinfo{person}{Tom Vercauteren}, \bibinfo{person}{Sebastien Ourselin},
  {and} \bibinfo{person}{M Jorge~Cardoso}.} \bibinfo{year}{2017}\natexlab{}.
\newblock \showarticletitle{Generalised dice overlap as a deep learning loss
  function for highly unbalanced segmentations}. In
  \bibinfo{booktitle}{\emph{Deep Learning in Medical Image Analysis and
  Multimodal Learning for Clinical Decision Support: Third International
  Workshop, DLMIA 2017, and 7th International Workshop, ML-CDS 2017, Held in
  Conjunction with MICCAI 2017, Qu{\'e}bec City, QC, Canada, September 14,
  Proceedings 3}}. Springer, \bibinfo{pages}{240--248}.
\newblock


\bibitem[Tajbakhsh et~al\mbox{.}(2015)]%
        {tajbakhsh2015automated}
\bibfield{author}{\bibinfo{person}{Nima Tajbakhsh},
  \bibinfo{person}{Suryakanth~R Gurudu}, {and} \bibinfo{person}{Jianming
  Liang}.} \bibinfo{year}{2015}\natexlab{}.
\newblock \showarticletitle{Automated polyp detection in colonoscopy videos
  using shape and context information}.
\newblock \bibinfo{journal}{\emph{IEEE transactions on medical imaging}}
  \bibinfo{volume}{35}, \bibinfo{number}{2} (\bibinfo{year}{2015}),
  \bibinfo{pages}{630--644}.
\newblock


\bibitem[Thambawita et~al\mbox{.}(2021a)]%
        {thambawita2021deepsynthbody}
\bibfield{author}{\bibinfo{person}{Vajira Thambawita},
  \bibinfo{person}{Steven~A Hicks}, \bibinfo{person}{Jonas Isaksen},
  \bibinfo{person}{Mette~Haug Stensen}, \bibinfo{person}{Trine~B Haugen},
  \bibinfo{person}{J{\O}rgen Kanters}, \bibinfo{person}{Sravanthi Parasa},
  \bibinfo{person}{Thomas de Lange}, \bibinfo{person}{H{\aa}vard~D Johansen},
  \bibinfo{person}{Dag Johansen}, {et~al\mbox{.}}}
  \bibinfo{year}{2021}\natexlab{a}.
\newblock \showarticletitle{DeepSynthBody: the beginning of the end for data
  deficiency in medicine}. In \bibinfo{booktitle}{\emph{2021 International
  Conference on Applied Artificial Intelligence (ICAPAI)}}. IEEE,
  \bibinfo{pages}{1--8}.
\newblock


\bibitem[Thambawita et~al\mbox{.}(2021b)]%
        {9462062}
\bibfield{author}{\bibinfo{person}{Vajira Thambawita},
  \bibinfo{person}{Steven~A. Hicks}, \bibinfo{person}{Jonas Isaksen},
  \bibinfo{person}{Mette~Haug Stensen}, \bibinfo{person}{Trine~B. Haugen},
  \bibinfo{person}{JØrgen Kanters}, \bibinfo{person}{Sravanthi Parasa},
  \bibinfo{person}{Thomas de Lange}, \bibinfo{person}{Håvard~D. Johansen},
  \bibinfo{person}{Dag Johansen}, \bibinfo{person}{Hugo~L. Hammer},
  \bibinfo{person}{Pål Halvorsen}, {and} \bibinfo{person}{Michael~A.
  Riegler}.} \bibinfo{year}{2021}\natexlab{b}.
\newblock \showarticletitle{DeepSynthBody: the beginning of the end for data
  deficiency in medicine}. In \bibinfo{booktitle}{\emph{2021 International
  Conference on Applied Artificial Intelligence (ICAPAI)}}.
  \bibinfo{pages}{1--8}.
\newblock
\urldef\tempurl%
\url{https://doi.org/10.1109/ICAPAI49758.2021.9462062}
\showDOI{\tempurl}


\bibitem[Thambawita et~al\mbox{.}(2021c)]%
        {thambawita2021deepfake}
\bibfield{author}{\bibinfo{person}{Vajira Thambawita}, \bibinfo{person}{Jonas~L
  Isaksen}, \bibinfo{person}{Steven~A Hicks}, \bibinfo{person}{Jonas Ghouse},
  \bibinfo{person}{Gustav Ahlberg}, \bibinfo{person}{Allan Linneberg},
  \bibinfo{person}{Niels Grarup}, \bibinfo{person}{Christina Ellervik},
  \bibinfo{person}{Morten~Salling Olesen}, \bibinfo{person}{Torben Hansen},
  {et~al\mbox{.}}} \bibinfo{year}{2021}\natexlab{c}.
\newblock \showarticletitle{DeepFake electrocardiograms using generative
  adversarial networks are the beginning of the end for privacy issues in
  medicine}.
\newblock \bibinfo{journal}{\emph{Scientific reports}} \bibinfo{volume}{11},
  \bibinfo{number}{1} (\bibinfo{year}{2021}), \bibinfo{pages}{21896}.
\newblock


\bibitem[Thambawita et~al\mbox{.}(2020)]%
        {thambawita2020extensive}
\bibfield{author}{\bibinfo{person}{Vajira Thambawita}, \bibinfo{person}{Debesh
  Jha}, \bibinfo{person}{Hugo~Lewi Hammer}, \bibinfo{person}{H{\aa}vard~D
  Johansen}, \bibinfo{person}{Dag Johansen}, \bibinfo{person}{P{\aa}l
  Halvorsen}, {and} \bibinfo{person}{Michael~A Riegler}.}
  \bibinfo{year}{2020}\natexlab{}.
\newblock \showarticletitle{An extensive study on cross-dataset bias and
  evaluation metrics interpretation for machine learning applied to
  gastrointestinal tract abnormality classification}.
\newblock \bibinfo{journal}{\emph{ACM Transactions on Computing for
  Healthcare}} \bibinfo{volume}{1}, \bibinfo{number}{3} (\bibinfo{year}{2020}),
  \bibinfo{pages}{1--29}.
\newblock


\bibitem[Thambawita et~al\mbox{.}(2022)]%
        {thambawita2022singan}
\bibfield{author}{\bibinfo{person}{Vajira Thambawita}, \bibinfo{person}{Pegah
  Salehi}, \bibinfo{person}{Sajad~Amouei Sheshkal}, \bibinfo{person}{Steven~A
  Hicks}, \bibinfo{person}{Hugo~L Hammer}, \bibinfo{person}{Sravanthi Parasa},
  \bibinfo{person}{Thomas~de Lange}, \bibinfo{person}{P{\aa}l Halvorsen}, {and}
  \bibinfo{person}{Michael~A Riegler}.} \bibinfo{year}{2022}\natexlab{}.
\newblock \showarticletitle{SinGAN-Seg: Synthetic training data generation for
  medical image segmentation}.
\newblock \bibinfo{journal}{\emph{PloS one}} \bibinfo{volume}{17},
  \bibinfo{number}{5} (\bibinfo{year}{2022}), \bibinfo{pages}{e0267976}.
\newblock


\bibitem[Thambawita et~al\mbox{.}(2021d)]%
        {thambawita2021id}
\bibfield{author}{\bibinfo{person}{Vajira~L Thambawita}, \bibinfo{person}{Inga
  Str{\"u}mke}, \bibinfo{person}{Steven Hicks}, \bibinfo{person}{Michael~A
  Riegler}, \bibinfo{person}{P{\aa}l Halvorsen}, {and}
  \bibinfo{person}{Sravanthi Parasa}.} \bibinfo{year}{2021}\natexlab{d}.
\newblock \showarticletitle{ID: 3523524 Data augmentation using generative
  adversarial networks for creating realistic artificial colon polyp images:
  validation study by endoscopists}.
\newblock \bibinfo{journal}{\emph{Gastrointestinal Endoscopy}}
  \bibinfo{volume}{93}, \bibinfo{number}{6} (\bibinfo{year}{2021}),
  \bibinfo{pages}{AB190}.
\newblock


\bibitem[Vinsard et~al\mbox{.}(2019)]%
        {vinsard2019quality}
\bibfield{author}{\bibinfo{person}{Daniela~Guerrero Vinsard},
  \bibinfo{person}{Yuichi Mori}, \bibinfo{person}{Masashi Misawa},
  \bibinfo{person}{Shin-ei Kudo}, \bibinfo{person}{Amit Rastogi},
  \bibinfo{person}{Ulas Bagci}, \bibinfo{person}{Douglas~K Rex}, {and}
  \bibinfo{person}{Michael~B Wallace}.} \bibinfo{year}{2019}\natexlab{}.
\newblock \showarticletitle{Quality assurance of computer-aided detection and
  diagnosis in colonoscopy}.
\newblock \bibinfo{journal}{\emph{Gastrointestinal endoscopy}}
  \bibinfo{volume}{90}, \bibinfo{number}{1} (\bibinfo{year}{2019}),
  \bibinfo{pages}{55--63}.
\newblock


\bibitem[Zhou et~al\mbox{.}(2018)]%
        {zhou2018unet++}
\bibfield{author}{\bibinfo{person}{Zongwei Zhou}, \bibinfo{person}{Md~Mahfuzur
  Rahman~Siddiquee}, \bibinfo{person}{Nima Tajbakhsh}, {and}
  \bibinfo{person}{Jianming Liang}.} \bibinfo{year}{2018}\natexlab{}.
\newblock \showarticletitle{Unet++: A nested u-net architecture for medical
  image segmentation}. In \bibinfo{booktitle}{\emph{Deep Learning in Medical
  Image Analysis and Multimodal Learning for Clinical Decision Support: 4th
  International Workshop, DLMIA 2018, and 8th International Workshop, ML-CDS
  2018, Held in Conjunction with MICCAI 2018, Granada, Spain, September 20,
  2018, Proceedings 4}}. Springer, \bibinfo{pages}{3--11}.
\newblock


\end{thebibliography}


\end{document}